\documentclass[sigconf,screen]{acmart}
\usepackage{microtype}
\usepackage{adjustbox}
\usepackage{seqsplit}
\usepackage{ifthen}
\usepackage{multirow}
\usepackage{tikz}
\usepackage{fontawesome5}
\usetikzlibrary{arrows.meta,positioning,calc}

\AtBeginDocument{%
  \providecommand\BibTeX{{%
    \normalfont B\kern-0.5em{\scshape i\kern-0.25em b}\kern-0.8em\TeX}}}

\copyrightyear{2026}
\acmYear{2026}
\setcopyright{cc}
\setcctype{by}
\acmDOI{10.1145/3832783.3834425}
\acmISBN{979-8-4007-2882-2/2026/10} 

\acmConference[ASE '26]{Proceedings of the 41st IEEE/ACM International Conference on Automated Software Engineering}{October 12--16, 2026}{Munich, Germany}
\acmBooktitle{Proceedings of the 41st IEEE/ACM International Conference on Automated Software Engineering (ASE '26), October 12--16, 2026, Munich, Germany}

\usepackage{tabularx}
\usepackage{booktabs}

\usepackage{xspace}

\usepackage{enumitem}

\usepackage[dvipsnames,svgnames,x11names]{xcolor}

\definecolor{keywordcolor}{HTML}{0000FF}
\definecolor{typecolor}{HTML}{007070}
\definecolor{functioncolor}{HTML}{7928a1}
\definecolor{stringcolor}{HTML}{A31515}
\definecolor{commentcolor}{HTML}{6A9955}
\definecolor{numbercolor}{HTML}{098658}

\usepackage{listings}
\usepackage[scaled]{beramono}

\usepackage{setspace,caption}
\usepackage{flushend} %

\usepackage{url} %

\usepackage{cleveref}

\newcommand{\todo}[1]{{\color{red}\bfseries [[#1]]}}
\ifdefined\notodocomments
  \renewcommand{\todo}[1]{\relax}
\fi

\newif\ifanonymous
\anonymoustrue

\newcommand{\anonurl}[1]{\ifanonymous URL removed for anonymity.\else\url{#1}\fi}

\def\|#1|{\mathid{#1}}
\newcommand{\mathid}[1]{\ensuremath{\mathit{#1}}}
\def\<#1>{\codeid{#1}}
\protected\def\codeid#1{\ifmmode{\mbox{\smaller\ttfamily{#1}}}\else{\smaller\ttfamily #1}\fi}
\newlist{researchquestions}{enumerate}{1}
\setlist[researchquestions]{label*=\textbf{RQ\arabic*}}

\newcommand{\ie}{i.e.,\xspace}
\newcommand{\eg}{e.g.,\xspace}
\newcommand{\etal}{et al.\xspace}

\newboolean{showcomments}
\setboolean{showcomments}{true}

\ifthenelse{\boolean{showcomments}}
{\newcommand{\nb}[2]{
		\fbox{\bfseries\sffamily\scriptsize#1}
		{\sf\small$\blacktriangleright$\textit{#2}$\blacktriangleleft$}
	}
	
}
{\newcommand{\nb}[2]{}
	
}

\setlist[itemize]{leftmargin=*, noitemsep, topsep=1pt}
\setlist[enumerate]{leftmargin=*, noitemsep, topsep=1pt}

\begin{document}

\title{On the Reliability of Code Comprehension Proxies}

\author{Erfan Arvan}
\affiliation{
  \institution{New Jersey Institute of Technology}
  \city{Newark}
  \state{New Jersey}
  \country{USA}
}
\email{ea442@njit.edu}

\author{Nadeeshan De Silva}
\affiliation{
  \institution{William \& Mary}
  \city{Williamsburg}
  \state{Virginia}
  \country{USA}
}
\email{kgdesilva@wm.edu}

\author{Oscar Chaparro}
\affiliation{
  \institution{William \& Mary}
  \city{Williamsburg}
  \state{Virginia}
  \country{USA}
}
\email{oscarch@wm.edu}

\author{Martin Kellogg}
\affiliation{
  \institution{New Jersey Institute of Technology}
  \city{Newark}
  \state{New Jersey}
  \country{USA}
}
\email{martin.kellogg@njit.edu}

\renewcommand{\shortauthors}{Arvan, De Silva, Chaparro, and Kellogg}

\begin{abstract}
Prior work on code comprehension uses different \emph{comprehension proxies}---for example,
Likert-scale ratings or answers to input-output questions about program snippets, usually collected from students, to
approximate whether code is comprehensible to software engineers, but the relative reliability of these proxies
is not known.

This paper investigates the relative reliability of a collection of proxies common in the extant literature
with a pair of human studies.
First, we conducted an expert-consensus study with a panel of five professional software engineers to establish a
ground-truth comprehensibility ranking of eight code snippets by adapting the Delphi expert-consensus protocol.
The Delphi protocol is widely used for expert consensus under conditions of uncertainty in other domains such as
medicine and national-security forecasting, but to our knowledge, this is its first application to code comprehension research.
Second, we conducted a study with 44 student participants who completed comprehension tasks, allowing us to measure 14
comprehension proxies derived from the literature on the same set of eight code snippets.
Finally, we conducted a correlation analysis on the results, concluding that proxies
1) derived from input-output questions and
2) that measure response time rather than accuracy are especially reliable.
We also found that proxies derived from questions about program syntax (rather than semantics)
are especially unreliable, regardless of measurement strategy, which draws into question the reliability of parts of the existing
comprehensibility literature.
\looseness=-1

\label{dummy-label-for-etags:abstract}

\end{abstract}

\ccsdesc[500]{Human-centered computing~Empirical studies in HCI}
\ccsdesc[300]{General and reference~Empirical studies}
\ccsdesc[100]{Software and its engineering~Maintaining software}

\keywords{code comprehension, comprehensibility proxies, Delphi method, controlled human study}

\maketitle

\section{Introduction}
\label{sec:intro}

Code comprehension---the process of understanding source code---is a fundamental software engineering activity, yet there is little agreement on how to measure it. Most maintenance tasks, including refactoring, debugging, and extending functionality, depend on a developer’s ability to build an accurate mental model of the code. Prior studies estimate that developers spend between 58\% and 70\% of their time on code comprehension~\cite{percent1,scalabrino2017automatically}.

Comprehension is an internal cognitive process that cannot be directly observed, so most prior empirical studies rely on \emph{comprehension proxies}:
observable \textit{measurements} collected from humans performing specific comprehension \textit{tasks}. Examples of such tasks include summarizing code~\cite{S4, S11, S24, S35, 33-7, 33-8, S38, S42, S49} and fixing bugs~\cite{S38, S55, S64, S87}. Common measurements include subjective ratings of code understandability~\cite{S4, S34, 33-7, S42, S49, 33-20}, the time required to answer questions about the code~\cite{Scalabrino:TSE19, S11, S39, S48, S50, S57, S59, S60, 33-24}, and the correctness of those answers~\cite{Scalabrino:TSE19, S11, NovicesVSprof1:BIBE2016, S48, S50, S57, S59, S60, 33-24}.
These proxies are intended to capture the difficulty of comprehension (\ie ``code comprehensibility'').
For example, longer response times %
are assumed to indicate that a code snippet is harder for humans to understand.
Comprehension proxies have been used not only to investigate how different factors influence comprehensibility~\cite{S4,S49,Peitek:TSE18,Siegmund:ICSE14,Stapleton:ICPC2020,Wyrich:ACS23} but also to develop predictive models that help developers measure and control comprehension difficulty during software evolution~\cite{Scalabrino:TSE19,Raymond:TSE10,Lavazza:ESE2023, De:2025,DeSilva2026EASE}.
More recently, with advancements in large language models and automated AI-based tools, 
comprehension proxies have also been used to evaluate whether such tools generate human-understandable code 
(e.g., UTGen~\cite{deljouyi2025leveraging}).

While these proxies are widely used in code comprehensibility studies, their reliability remains largely unexamined.

This creates a fundamental problem: it is unclear which proxies, if any, accurately capture the underlying construct of code comprehension difficulty.
Further complicating the problem, this underlying construct is not well-defined, and various studies operationalize it in different ways ~\cite{Wyrich:ACS23}. As a result, findings across the literature can be difficult to compare, and conclusions drawn from these proxies may be misleading.

In this work, we address this problem by defining code comprehensibility in terms of expert agreement about comprehension difficulty. In the absence of an objective definition, we argue that consistent agreement among experienced software engineers provides a practical and reliable approximation of the construct. Intuitively, if experts consistently agree that certain code is easier or harder to understand, this agreement can serve as a reference point against which proxies can be evaluated.

To operationalize this idea, we conducted two complementary studies on eight shared code snippets. %
First, we carried out an \textbf{expert study} with five senior software engineers using the Delphi expert-consensus protocol~\cite{delphi,delphiOriginal,randManual}, a structured method for eliciting and refining expert judgment under uncertainty. Over multiple rounds, participants ranked the snippets by comprehension difficulty and iteratively resolved disagreements through written feedback and discussion. The Delphi protocol has been successfully applied in medicine~\cite{m1,m2,m3,m4}, geopolitical forecasting~\cite{g1,g2}, and other domains~\cite{o1, o3,o4,o5}. 
Second, we conducted a \textbf{student study} with 44 computer science (CS) undergraduates from two U.S. universities. The students completed a set of tasks derived from commonly used comprehension proxies in the literature, including time-based, accuracy-based, and subjective measures. 
We used students for this part of the study because approximately 90\% of studies in a recent survey of code comprehension research measured their proxies with student participants~\cite{Wyrich:ACS23}.
We then conducted \textbf{a correlation analysis} where we analyzed how well each proxy correlates with the expert-consensus ranking, treating the latter as a ground-truth approximation of code comprehensibility.
\looseness=-1

Our results reveal three main findings. First,  proxies based on \textbf{input-output reasoning} about program behavior, including both response time and answer correctness,  show the strongest relationship with expert judgments. Second, \textbf{time-based measures} consistently outperform direct correctness-based measures; the time required to answer input-output questions was the single best-performing proxy.
These results imply that future studies of code comprehensibility can confidently approximate comprehensibility with time measurements of input-output questions.
Third, several commonly used proxies, including \textbf{subjective Likert-scale ratings and human-judged free-text code summaries}, show weak correlations with expert consensus. Notably, \textbf{proxies based on syntactic questions} (\eg identifying structural properties of the code) exhibit near-zero or negative correlations, suggesting that they may misrepresent comprehension effort.
As a result, we can reinterpret some results from the literature that solely rely on syntax-derived proxies (\eg~\cite{33-4,33-29}).

In summary, our contributions are:
\begin{itemize}
	\item A conceptualization of code comprehensibility based on expert agreement about comprehension difficulty, enabling the construction of ground-truth data (\cref{sec:overview}).
	\item An expert study using the Delphi protocol to construct a consensus-based ground truth (\cref{sec:expert-study}).
	\item A student study that collects 14 comprehension proxies commonly used in prior work (\cref{sec:student-study}).
	\item A correlation analysis comparing these proxies against expert-consensus judgments, showing that time-based and input-output proxies align more closely with expert assessments than other proxies (\cref{sec:results}).
	\item A discussion of the implications of our findings for interpreting prior code comprehension research (\cref{sec:discussion}).
\end{itemize}

\section{Background and Related Work}
\label{sec:relatedwork}

\textbf{Universal definition for code comprehensibility.} Despite its importance, the code comprehension construct still lacks a precise definition that primary studies can rely on in their design, let alone a standardized measurement method or even consensus on which features to include in such measurements~\cite{chin2026put, Wyrich:ACS23}.
As a consequence, what is measured is often implicitly defined by the method of measurement itself (\ie the comprehension proxy). %
A comprehension proxy has two parts: a task and a measure.
Common tasks in the literature include reading code~\cite{Karas:TSE2024}, answering comprehension questions~\cite{Park:ESE24, Peitek:ICPC20, Abdelsalam:ESE2026, Gao:EASE2025}, summarizing functionality~\cite{Karas:TSE2024,Peitek:ICPC20, Flint:2026}, or finding or fixing a bug~\cite{Nielebock:ESE2019, Flint:2026}.
Common measures in the literature are time~\cite{Nielebock:ESE2019, Peitek:ICPC20, Peitek:ICSE2021, Karas:TSE2024, Park:ESE24, Alakmeh:ASE2024, Elfares:2025, Abdelsalam:ESE2026}, accuracy~\cite{Nielebock:ESE2019, Peitek:ICPC20, Peitek:ICSE2021, Ribeiro:ESE23, Karas:TSE2024, Park:ESE24, Alakmeh:ASE2024, Elfares:2025, Flint:2026, Abdelsalam:ESE2026}, eye movement~\cite{Peitek:ICPC20, Abbad-Andaloussi:ICPC22, Park:ESE24, Elfares:2025, Abdelsalam:ESE2026, Flint:2026}, or subjective ratings~\cite{Peitek:ICSE2021, Ribeiro:ESE23, Alakmeh:ASE2024, Elfares:2025, Bergum:TOSEM2026, Abdelsalam:ESE2026}.
The task and measure are inseparable when interpreting proxy measurements. For example, accuracy may be meaningless if the task is poorly designed (\eg due to vague or trivial questions),
but it can be informative when paired with a well-designed task. Likewise, a meaningful task may yield little insight when paired with an uninformative measure.

\textbf{Generalization from students to developers.}
The problem is further complicated by participant selection. Based on data reported by Wyrich \etal~\cite{Wyrich:ACS23}, approximately 90\% of code comprehensibility studies between 2010 and 2019 included students as participants~\cite{S4, scalabrino2017automatically, chen2017evaluating, Fakhoury:2018ICPC, Scalabrino:TSE19}, and this trend has continued in more recent years~\cite{Nielebock:ESE2019, Peitek:ICSE2021, Ribeiro:ESE23, Sampaio:ICCQ2024, Karas:TSE2024, Abdelsalam:ESE2026}. These studies often recruit students because they are inexpensive and readily available, yet generalize the results to professional software engineers, who are the intended target population.

We regard the assumption that results obtained from students can be unconditionally generalized to professionals as a methodological flaw because code comprehension is a skill shaped by professional experience.
Moreover, multiple studies confirm that professionals and novices differ substantially in how they approach code comprehension~\cite{Aljehane:HCI2023, novicesVSprof2:ESE2002, NovicesVSprof1:BIBE2016, 33-8}. At the neurological level, the two groups activate different brain regions when reading code~\cite{NovicesVSprof1:BIBE2016}, suggesting that expertise fundamentally reshapes how the brain processes code. 
These differences also manifest in reading behavior: eye-tracking data reveal that experts do not read code in the order it is written, instead navigating non-linearly toward semantically relevant regions, whereas novices tend to treat source code as a narrative, reading it sequentially~\cite{33-8}. Accordingly, experts concentrate their attention on functional units and object interactions that serve program goals, while novices focus on structural units and literal statements~\cite{novicesVSprof2:ESE2002}. 
This targeted behavior is further reflected in visual attention patterns: experts efficiently direct their gaze toward information relevant to an area of interest, whereas novices are more easily distracted by irrelevant content~\cite{Aljehane:HCI2023, 33-8}.
These findings motivate our study by suggesting that proxies derived from prior studies with students may not accurately capture the comprehension processes of professionals, and so may not serve as reliable indicators of code comprehensibility.

\subsection{Comprehension Tasks in Prior Work}
\label{sec:rw-tasks}

\begin{table}
\centering
\caption{Taxonomy of comprehension tasks based on Wyrich~\etal~\cite{Wyrich:ACS23}, with updated study counts through March 2026.}
\resizebox{\columnwidth}{!}{
\begin{tabular}{p{3cm} l c}
  \toprule
 \textbf{Task Category} & \textbf{Task} & \textbf{\# of Studies} \\ 
  \midrule

\multirow{6}{3cm}{Tc1: Provide information about the code}
    & Answer comprehension questions & 37 \\ 
    & Determine output of a program & 30 \\ 
    & Summarize code verbally or textually & 25 \\ 
    & Recall (parts of) the code & 7 \\ 
    & Match with diagram or similar code & 3 \\
    & Determine code trace & 1 \\ 
  \midrule

\multirow{6}{3cm}{Tc2: Provide personal opinion}
  & Rate code comprehensibility & 13 \\ 
  & Other subjective indications & 7 \\ 
  & Rate task difficulty or task load & 5 \\ 
  & Rate confidence in answer/understanding & 5 \\ 
  & Compare or rank code snippets & 5 \\
  & Rate own code understanding & 2 \\  
  \midrule

\multirow{2}{3cm}{Tc3: Debug code}
  & Find a bug & 9 \\ 
  & Extend the code & 4 \\ 
  \midrule

\multirow{4}{3cm}{Tc4: Maintain code}
    & Fix a bug & 4 \\ 
    & Cloze test on code & 3 \\ 
    & Refactor code & 3 \\ 
    & Determine if code is correct & 1 \\
 
   \bottomrule
\end{tabular}
}
\label{tab:proxies-wyrich}
\end{table}

\Cref{tab:proxies-wyrich} extends the taxonomy of comprehension tasks in Wyrich~\etal's survey~\cite{Wyrich:ACS23}
with papers published on or before March 25, 2026 but after their literature search.
While there is significant diversity in tasks, 
the literature has a strong bias toward \textbf{Tc1} and, to a lesser extent, 
\textbf{Tc2}. The most common tasks in these categories are answering questions of 
various types, determining program output given some inputs, free-text summarization, and subjective ratings; together, these 
tasks dominate the literature. %

\textbf{Open coding for comprehensibility questions.}
\label{sec:open-coding}
The task ``answer comprehension questions'' is highly abstract: one can conceive of questions 
targeting many different aspects of code---semantics, syntax, functionality, and so on.
So, we performed an open-coding analysis ourselves on the questions actually asked in the relevant papers~\cite{Scalabrino:TSE19,33-2,33-3,33-4,33-5,S34,33-7,33-8,33-9,penningtonModel, 33-11,33-12,33-13,33-14,33-15,33-16,33-17,33-18,33-19,33-20,33-21,33-22,33-23,33-24,33-25,33-26,33-27,33-28,33-29,33-30,33-31,33-32,33-33}. 

We first extracted all the questions. Two authors then 
independently performed open coding to categorize these questions. Following 
this initial phase, the authors discussed their labels and converged on a unified 
taxonomy of question categories. To assess inter-rater agreement, each author's 
initial labels were mapped to the final taxonomy and Cohen's kappa~\cite{cohen} was computed, 
yielding $\kappa = 0.89$. The resulting taxonomy, simplified for space, appears in \cref{fig:taxonomy_text}.
It guided the choice of questions in our own study (\cref{sec:question-design}).
Our replication package has a full list of questions and their categorization~\cite{replication-package}.

\begin{figure}[t]
\centering
\scriptsize
\begin{minipage}{0.95\linewidth}
\begin{verbatim}
Question categories (111)
|-- Program Function (27)
|   (e.g., What does the identifier "ListLimit" signify?)
|   
|-- Semantics (49)
|   |-- Program Analysis (15)
|   |   (e.g., Does X affect Y?)
|   |
|   `-- Simulated Execution (34)
|       |-- Conditional (7)
|       |   (e.g., Is an error message printed?)
|       |
|       |-- Output (10)
|       |   (e.g., What output is produced for a given input?)
|       |
|       `-- Internal State (17)
|           (e.g., State the final value of "X", given an input "Y"?)
|          
|-- Syntax (30)
|   |-- Global (18)
|   |   (e.g., How many loops are there?)
|   |
|   `-- Local (12)
|       (e.g., Is variable "Z" initialized to zero?)
|
`-- Other (5)
\end{verbatim}
\end{minipage}
\caption{Simplified taxonomy of question categories derived from prior studies, with the number of instances per category and one representative example per leaf.
The full categorization has more levels of detail and is available in our replication package~\cite{replication-package}.
}
\label{fig:taxonomy_text}
\end{figure}

\subsection{Comprehension Measures in Prior Work}
\label{sec:rw-measures}

\begin{table}[t]
\centering
\small
\renewcommand{\arraystretch}{1.1}
\caption{Frequency of comprehension measure types based on the taxonomy of Wyrich~\etal~\cite{Wyrich:ACS23}, with updated study counts through March 2026.}
\begin{tabular}{lclc}
\toprule
\textbf{Measure Type} & \textbf{Studies} & \textbf{Measure Type} & \textbf{Studies} \\
\midrule
Correctness & 80 & Physiological & 30 \\
Time & 50 & Aggregation & 6 \\
Subjective rating & 37 & Others & 6 \\
\bottomrule
\end{tabular}
\label{tab:measures}
\end{table}

\Cref{tab:measures} extends the comprehension measure survey data from Wyrich~\etal~\cite{Wyrich:ACS23} with
that from studies published before March 25, 2026.
\emph{Correctness} (also referred to as \emph{accuracy} ~\cite{S29}) compares a study participant's (\ie a subject) answer to a known ground truth~\cite{Nielebock:ESE2019, Peitek:ICPC20, Peitek:ICSE2021, Ribeiro:ESE23, Karas:TSE2024, Park:ESE24, Alakmeh:ASE2024, Elfares:2025, Flint:2026, Abdelsalam:ESE2026}.
\emph{Time} captures how long a subject takes to complete 
a task, regardless of answer correctness~\cite{Nielebock:ESE2019, Peitek:ICPC20, Peitek:ICSE2021, Karas:TSE2024, Park:ESE24, Alakmeh:ASE2024, Elfares:2025, Abdelsalam:ESE2026}.
\emph{Subjective ratings} are self-reported perceptions, such as 
Likert-scale responses to questions~\cite{Nielebock:ESE2019, Peitek:ICSE2021, Ribeiro:ESE23, Abdelsalam:ESE2026, Alakmeh:ASE2024, Elfares:2025, Bergum:TOSEM2026}.
\emph{Physiological} measurements capture responses of the human body or brain when performing code comprehensibility tasks. 
Eye-tracking~\cite{Peitek:ICPC20, Abbad-Andaloussi:ICPC22, Park:ESE24, Abdelsalam:ESE2026}, fMRI~\cite{Peitek:ICSE2021} and EEG~\cite{Bergum:TOSEM2026} are the most common measures in this category.
\emph{Aggregation} combines two or more other measures~\cite{Ribeiro:ESE23, Alakmeh:ASE2024, Gao:EASE2025}; 
for example, ``time to correctly answer a question'' aggregates correctness 
and time measures.
\emph{Other} includes measures that do not fit into the above categories, such as using code metrics to measure comprehensibility~\cite{33-8, Nielebock:ESE2019}.
This taxonomy guided our choice of measures in our study (\cref{sec:our-measures}).

\section{Methodology Overview}
\label{sec:overview}

\begin{figure}
  \centering
  \begin{adjustbox}{max width=\columnwidth, max height=0.42\textheight}
    \begin{tikzpicture}[
      >={Stealth[length=3mm,width=2mm]},
      flow/.style={
        -{Stealth[length=3mm,width=2mm]},
        draw=black!70,
        line width=0.9pt,
        rounded corners=2pt
      },
      childflow/.style={
        -{Stealth[length=2.2mm,width=1.5mm]},
        draw=black!55,
        line width=0.7pt
      },
      plainline/.style={
        draw=black!70,
        line width=0.9pt,
        rounded corners=2pt
      },
      conceptualflow/.style={
        -{Stealth[length=3mm,width=2mm]},
        draw=violet!80!black,
        dashed,
        line width=1pt,
        rounded corners=3pt
      },
      bluebox/.style={
        rounded corners=6pt,
        draw=blue!75!black,
        fill=blue!3,
        line width=1.1pt
      },
      tealbox/.style={
        rounded corners=6pt,
        draw=teal!75!black,
        fill=teal!3,
        line width=1.1pt
      },
      orangebox/.style={
        rounded corners=6pt,
        draw=orange!70!black,
        fill=orange!4,
        line width=1.1pt
      },
      conceptbox/.style={
        rounded corners=6pt,
        draw=violet!80!black,
        fill=violet!2,
        dashed,
        line width=1.1pt
      },
      tealchildbox/.style={
        rounded corners=4pt,
        draw=teal!55!black,
        fill=teal!6,
        line width=0.8pt
      },
      orangechildbox/.style={
        rounded corners=4pt,
        draw=orange!55!black,
        fill=orange!6,
        line width=0.8pt
      },
      bluetitle/.style={
        font=\rmfamily\bfseries\large,
        text=blue!70!black
      },
      tealtitle/.style={
        font=\rmfamily\bfseries\large,
        text=teal!70!black
      },
      orangetitle/.style={
        font=\rmfamily\bfseries\large,
        text=orange!60!black
      },
      purpletitle/.style={
        font=\rmfamily\bfseries\large,
        text=violet!75!black
      },
      subtitleteal/.style={
        font=\rmfamily\bfseries\normalsize,
        text=teal!60!black
      },
      subtitleorange/.style={
        font=\rmfamily\bfseries\normalsize,
        text=orange!55!black
      },
      body/.style={
        font=\rmfamily\normalsize,
        text=black!90
      },
      compactbody/.style={
        font=\rmfamily\normalsize,
        text=black!90
      },
      iconblue/.style={
        font=\fontsize{20}{20}\selectfont,
        text=blue!70!black
      },
      iconteal/.style={
        font=\fontsize{16}{16}\selectfont,
        text=teal!70!black
      },
      iconorange/.style={
        font=\fontsize{16}{16}\selectfont,
        text=orange!60!black
      },
      iconpurple/.style={
        font=\fontsize{19}{19}\selectfont,
        text=violet!75!black
      }
    ]

      \path[use as bounding box] (-6.0,-6.0) rectangle (6.0,8.7);

      \node[
        conceptbox,
        minimum width=5.2cm,
        minimum height=2.4cm
      ] (construct) at (-3.0,7.2) {};

      \node[iconpurple, anchor=west]
        at ([xshift=0.35cm,yshift=-0.55cm]construct.north west)
        {\faBullseye};

      \node[
        purpletitle,
        anchor=north west,
        text width=3.4cm,
        align=left
      ] at ([xshift=1.35cm,yshift=-0.40cm]construct.north west)
        {Population-Level Comprehensibility};

      \node[
        compactbody,
        anchor=north west,
        text width=4.4cm,
        align=left
      ] at ([xshift=0.35cm,yshift=-1.55cm]construct.north west)
        {- Comparative, idealized, unobservable construct (\cref{sec:ground-truth})};

      \node[
        bluebox,
        minimum width=5.2cm,
        minimum height=2.4cm
      ] (snippets) at (3.0,7.2) {};

      \node[iconblue, anchor=west]
        at ([xshift=0.35cm,yshift=-0.55cm]snippets.north west)
        {\faCode};

      \node[
        bluetitle,
        anchor=north west,
        text width=3.4cm,
        align=left
      ] at ([xshift=1.35cm,yshift=-0.40cm]snippets.north west)
        {Snippet Selection};

      \node[
        compactbody,
        anchor=north west,
        text width=4.4cm,
        align=left
      ] at ([xshift=0.35cm,yshift=-1.15cm]snippets.north west)
        {- 8 self-contained Java methods (\cref{sec:snippets})};

      \node[
        tealbox,
        minimum width=5.2cm,
        minimum height=6.7cm
      ] (expert-parent) at (-3.0,1.35) {};

      \node[iconteal, anchor=west]
        at ([xshift=0.35cm,yshift=-0.50cm]expert-parent.north west)
        {\faUsers};

      \node[
        tealtitle,
        anchor=north west,
        text width=3.4cm,
        align=left
      ] at ([xshift=1.35cm,yshift=-0.40cm]expert-parent.north west)
        {Expert Study};

      \draw[draw=teal!40, line width=0.6pt]
        ([xshift=0.30cm,yshift=-0.90cm]expert-parent.north west) --
        ([xshift=-0.30cm,yshift=-0.90cm]expert-parent.north east);

      \node[
        tealchildbox,
        anchor=north west,
        minimum width=4.6cm,
        minimum height=2.75cm
      ] (expert-method) at ([xshift=0.30cm,yshift=-1.10cm]expert-parent.north west) {};

      \node[iconteal, anchor=west]
        at ([xshift=0.20cm,yshift=-0.50cm]expert-method.north west)
        {\faCogs};

      \node[
        subtitleteal,
        anchor=north west,
        text width=2.7cm,
        align=left
      ] at ([xshift=1.20cm,yshift=-0.35cm]expert-method.north west)
        {Methodology};

      \node[
        compactbody,
        anchor=north west,
        text width=4.0cm,
        align=left
      ] at ([xshift=0.20cm,yshift=-1.20cm]expert-method.north west)
        {- 5 senior SEs\\
         - 3-round Delphi protocol\\
         - \cref{sec:expert-study}};

      \node[
        tealchildbox,
        anchor=north west,
        minimum width=4.6cm,
        minimum height=2.05cm
      ] (expert-rank) at ([xshift=0.30cm,yshift=-4.35cm]expert-parent.north west) {};

      \node[iconteal, anchor=west]
        at ([xshift=0.20cm,yshift=-0.50cm]expert-rank.north west)
        {\faListOl};

      \node[
        subtitleteal,
        anchor=north west,
        text width=2.7cm,
        align=left
      ] at ([xshift=1.20cm,yshift=-0.35cm]expert-rank.north west)
        {Expert Ranking};

      \node[
        compactbody,
        anchor=north west,
        text width=4.0cm,
        align=left
      ] at ([xshift=0.20cm,yshift=-1.15cm]expert-rank.north west)
                {- ground-truth ranking};

      \draw[childflow]
        (expert-method.south) -- (expert-rank.north);

      \node[
        orangebox,
        minimum width=5.2cm,
        minimum height=6.7cm
      ] (student-parent) at (3.0,1.35) {};

      \node[iconorange, anchor=west]
        at ([xshift=0.35cm,yshift=-0.50cm]student-parent.north west)
        {\faGraduationCap};

      \node[
        orangetitle,
        anchor=north west,
        text width=3.4cm,
        align=left
      ] at ([xshift=1.35cm,yshift=-0.40cm]student-parent.north west)
        {Student Study};

      \draw[draw=orange!40, line width=0.6pt]
        ([xshift=0.30cm,yshift=-0.90cm]student-parent.north west) --
        ([xshift=-0.30cm,yshift=-0.90cm]student-parent.north east);

      \node[
        orangechildbox,
        anchor=north west,
        minimum width=4.6cm,
        minimum height=2.75cm
      ] (student-method) at ([xshift=0.30cm,yshift=-1.10cm]student-parent.north west) {};

      \node[iconorange, anchor=west]
        at ([xshift=0.20cm,yshift=-0.50cm]student-method.north west)
        {\faCogs};

      \node[
        subtitleorange,
        anchor=north west,
        text width=2.7cm,
        align=left
      ] at ([xshift=1.20cm,yshift=-0.35cm]student-method.north west)
        {Methodology};

      \node[
        compactbody,
        anchor=north west,
        text width=4.0cm,
        align=left
      ] at ([xshift=0.20cm,yshift=-1.20cm]student-method.north west)
        {- 44 CS students\\
         - 14 proxies\\
         - \cref{sec:student-study}};

      \node[
        orangechildbox,
        anchor=north west,
        minimum width=4.6cm,
        minimum height=2.05cm
      ] (student-rank) at ([xshift=0.30cm,yshift=-4.35cm]student-parent.north west) {};

      \node[iconorange, anchor=west]
        at ([xshift=0.20cm,yshift=-0.50cm]student-rank.north west)
        {\faListOl};

      \node[
        subtitleorange,
        anchor=north west,
        text width=2.7cm,
        align=left
      ] at ([xshift=1.20cm,yshift=-0.35cm]student-rank.north west)
        {Proxy Ranking};

      \node[
        compactbody,
        anchor=north west,
        text width=4.0cm,
        align=left
      ] at ([xshift=0.20cm,yshift=-1.15cm]student-rank.north west)
        {- one ranking per proxy};

      \draw[childflow]
        (student-method.south) -- (student-rank.north);

      \node[
        bluebox,
        minimum width=5.2cm,
        minimum height=2.4cm
      ] (correlation) at (0,-4.5) {};

      \node[iconblue, anchor=west]
        at ([xshift=0.35cm,yshift=-0.55cm]correlation.north west)
        {\faChartLine};

      \node[
        bluetitle,
        anchor=north west,
        text width=3.4cm,
        align=left
      ] at ([xshift=1.35cm,yshift=-0.40cm]correlation.north west)
        {Correlation Analysis};

      \node[
        compactbody,
        anchor=north west,
        text width=4.4cm,
        align=left
      ] at ([xshift=0.35cm,yshift=-1.15cm]correlation.north west)
        {- Correlate rankings (\cref{sec:correlation})\\
         - Answer RQ1};

      \draw[conceptualflow]
        (construct.south) -- (expert-parent.north);

      \draw[flow]
        (snippets.south) -- (student-parent.north);

      \coordinate (snip-drop) at ($(snippets.south)+(0,-0.65)$);
      \draw[flow]
        (snippets.south) -- (snip-drop) -| ([xshift=0.9cm]expert-parent.north);

      \coordinate (merge-point) at ($(correlation.north)+(0,0.75)$);
      \draw[plainline]
        (expert-parent.south) |- (merge-point);
      \draw[plainline]
        (student-parent.south) |- (merge-point);
      \draw[flow]
        (merge-point) -- (correlation.north);

    \end{tikzpicture}
  \end{adjustbox}
  \caption{Overview of our methodology.}
  \label{fig:methodology-overview}
\end{figure}

Our goal is to evaluate whether proxies commonly used in prior studies
(typically measured with student participants) accurately reflect how
professional software engineers experience code comprehension
difficulty. This goal is important because students are likely to
remain a primary study population for practical reasons, such as cost
and availability, and the reliability of these proxies is unknown.
Our research question is:
\begin{researchquestions}
\item How well do common comprehension proxies from the literature correlate with expert-determined ground truth?
\end{researchquestions}
With this RQ, we are particularly interested in proxies that correlate very well (and
are therefore good candidates for future studies) or very poorly %
(and so prior work relying on them is suspect).
\looseness=-1

\Cref{fig:methodology-overview} summarizes our overall methodology.
We first define an idealized notion of code
comprehensibility (\cref{sec:ground-truth}) and then approximate it
empirically via the Delphi expert-consensus protocol (\cref{sec:delphi})
on a set of carefully-selected code snippets (\cref{sec:snippets}).
We then collect a range of proxies, drawn from
prior work, on these same snippets, using a student population similar to many
prior works (\cref{sec:our-proxies}). Finally, we analyze how well each
proxy collected from the student participants correlates with the expert-consensus judgments, treating the
latter as an approximation of ground-truth comprehension difficulty
(\cref{sec:correlation}).

Note that, in relying on expert judgments as this reference point and
collecting proxies from students, we are not claiming that experts and novices
comprehend code in the same way. Rather, our goal is to evaluate whether
proxies collected from student participants, who are commonly used in the
literature, approximate the comprehensibility judgments of professional
software engineers, the population such studies ultimately aim to generalize
to.

A tempting alternative approach to evaluate the proxies is to
conduct parallel studies with students and professionals and compare
proxy values directly between the two groups. However, this approach assumes
that at least one proxy is inherently valid---that is, that it can be
trusted to measure the underlying construct. In our setting, where no
proxy has been independently validated, such comparisons are
inconclusive: agreement between groups on a proxy does not establish
that the proxy itself is a reliable indicator of comprehension
difficulty.

Our overall methodology, including participant recruitment procedures,
questionnaires, comprehension tasks, and compensation methods for both
expert and student studies, was approved by the ethics review boards
of our respective institutions.

\subsection{Idealized Ground Truth}
\label{sec:ground-truth}

Comprehension is an internal cognitive
process and cannot be directly observed or measured. Moreover, comprehensibility
depends not only on the code itself, but also on its
human reader.
In this section, we explain how we define the ground truth of comprehensibility in the abstract
and how we approximate it in practice.

Prior work in psychology and program
comprehension~\cite{penningtonModel,f1,f2,f3,f4} characterizes
comprehension as the process of constructing a coherent internal
representation (or mental model) of the ``material'' (\eg text or
code), which can support tasks such as recall, reasoning, or
modification. We adopt this view and define code comprehensibility in
terms of how easily such a representation can be constructed relative
to a reference point: that is, one piece of code can be
easier or harder to understand than another, but assigning
a specific numerical value to “comprehensibility” is not defined.
At the level of an individual software engineer, we can thus %
model comprehensibility as a comparison between two code snippets: either one
is easier to understand than the other or
they are equally understandable. This captures the intuition that comprehension difficulty is better expressed comparatively than absolutely. This intuition aligns with the law of comparative judgment, which suggests greater consistency in relative than absolute judgments~\cite{Thurstone:LCJ2017}. Empirical evidence shows that absolute ratings suffer from heterogeneous scale interpretation, whereas comparative judgments yield more consistent and reliable measurements~\cite{Kinnear:BRM2025}. Based on relative code comparisons, we can establish a ranking of snippets by comprehensibility.

Building on this, we introduce an idealized notion of
\textit{population-level comprehensibility}, which reflects how a
broad population of software engineers would compare two
snippets. This construct is not directly observable, but serves as a
conceptual target: it represents the aggregate judgment we would
obtain if we could study all relevant developers under consistent
conditions.

Since such a population-level notion cannot be measured directly, we
approximate it using a group of experienced software engineers. We
refer to this approximation as \textit{expert-determined comprehensibility}: a
consensus-based ranking of code snippets produced by a panel of
experts. While this approximation is imperfect, it provides a
practical and reliable reference point for evaluating proxies,
particularly in the absence of an objective ground truth.

\subsection{Establishing Expert Consensus}
\label{sec:delphi}

For reliable expert consensus, we need a clear definition of expertise and a structured method for eliciting and refining judgments.
\looseness=-1

\subsubsection{Expert Definition}
We operationalize ``experts'' as professional software engineers with
``senior-level expertise'', demonstrated experience in real-world development projects, based on
professional progression and mentoring experience. Our recruitment criteria are proxies
for sustained exposure to diverse codebases and for an ability to
understand and assess how code is understood by others. More details
on expert recruitment are in \cref{sec:expert-participants}.

\subsubsection{Delphi Protocol}

We employed the Delphi method~\cite{delphi,delphiOriginal}, a structured approach for eliciting and refining expert judgments in settings where objective ground truth is unavailable \cite{m1,m2,m3,m4,g1,g2,o1, o3,o4,o5}. 
Our protocol consisted of three iterative rounds conducted on a custom
web platform, designed based on guidance published by Delphi experts~\cite {randManual}.
In each round, experts evaluated the same set of code
snippets by summarizing their functionality, assessing their
comprehensibility, and ranking them relative to one another. After
each round, participants received anonymized summaries of group
responses, reflected on disagreements and revised their
assessments. Anonymity was maintained throughout to reduce
bias. This process produced a stable, consensus-based ranking of
snippet comprehensibility, which we use as our approximation of ground
truth. Further details are provided in \cref{sec:expert-procedure}.

\begin{figure*}[t]
\centering
\small

\begin{minipage}[t]{0.48\textwidth}

\begin{lstlisting}
public static boolean isValidProjectName(String name) {
    if (name == null) return false;    
    if (name.startsWith(".")) return false;
    if ((name.length() < 1) || (name.length() > MAX_NAME_LENGTH)) {
        return false;
    }
    for (int i = 0; i < name.length(); i++) {
        char c = name.charAt(i);
        if (!Character.isLetterOrDigit(c) && !VALID_NAME_SET.contains(c)) {
            return false;
        }
    }
    return true;
}
\end{lstlisting}
\end{minipage}
\hfill
\begin{minipage}[t]{0.48\textwidth}

\begin{lstlisting}
public static float atan2(float y, float x) {
    float n = y / x;
    if (n != n)
        n = (y == x ? 1f : -1f);
    else if (n - n != n - n)
        x = 0f;
    if (x > 0)
        return atanUnchecked(n);
    else if (x < 0) {
        if (y >= 0) return atanUnchecked(n) + PI;
        return atanUnchecked(n) - PI;
    } else if (y > 0)
        return x + HALF_PI;
    else if (y < 0)
        return x - HALF_PI;
    return x + y;
}
\end{lstlisting}
\end{minipage}

\caption{Two example code snippets used in the study: \emph{isValidProjectName} (Snippet \#1) and \emph{atan2} (Snippet \#8).}
\label{fig:example_snippets}
\end{figure*}

\subsection{Snippet Selection}
\label{sec:snippets}
A key methodological decision in code comprehension studies is the
choice of code snippets. We use real-world code because our goal is to
draw conclusions relevant to professional software practice. At the
same time, the snippet set must support controlled comparison by
limiting confounding factors such as project-specific background
knowledge, code length, and code-formatting differences. Since the snippets from the prior studies did not meet these criteria, we decided to construct a new set of snippets for our study.  
We constructed the snippet set in three stages: project filtering,
method filtering, and redundancy reduction.
Figure~\ref{fig:example_snippets} shows two example
snippets; the other snippets are in our replication
package~\cite{replication-package}.

\subsubsection{Project Filtering}

We identified candidate open-source projects hosted in GitHub with SEART's repository search infrastructure~\cite{seart}.
To favor mature, active, and widely used open source projects, we
required candidate repositories to satisfy the following criteria: (1)
written in Java (because it is the most common language in prior
studies~\cite{Wyrich:ACS23} and is popular in industry~\cite{TIOBEIndex}),
(2) at least 1,000 commits, (3) at least 1,000 pull
requests, (4) at least 1,000 stars, and (5) at least one recent commit
within a three-week window after the data collection date (February 2,
2025).  These filters yielded 16 candidate projects (e.g.,
apache/kafka~\cite{kafka}, libgdx/libgdx~\cite{libgdx}). The full list
of projects is provided in our replication
package~\cite{replication-package}.

\subsubsection{Method Filtering}

Following prior work~\cite{Wyrich:ACS23}, we use individual methods as
the unit of analysis. Method-level comprehension is common in
practice, since developers often need to understand specific methods
in order to reuse, modify, or integrate them during development and
maintenance tasks.
We used these inclusion criteria to select methods:
1) all types in the signature and body must be from the JDK,
2) no annotations in signature, 3) static methods only,
4) must have a Javadoc comment, 5) must have at least one
parameter and return a value, and 6) must have 18-22 non-comment, non-blank
lines of code.
Criteria 1, 2, and 3 ensure the snippets are self-contained,
so that participants do not require project-specific knowledge.
Criteria 4 and 5 are necessary for our experimental design: Javadoc
comments are needed as ground-truth for free-text summary tasks,
and input/output question tasks require at least one input and at
least one output. Criterion 6 controls for size but keeps each task within a reasonable time budget while
still using snippets that implemented self-contained behavior.
Applying these criteria produced 21 candidate methods.

\subsubsection{Redundancy Reduction}

Thirteen of the 21 methods were identical or near-identical implementations 
appearing in multiple locations within the same project (\eg \emph{atan2},
\emph{atan2Deg}, and \emph{atan2Deg360} from libgdx~\cite{libgdx}).
We randomly retained one representative from each such group, 
yielding a final set of
eight methods. These eight were used for both the expert and
student studies.

\subsubsection{Code Presentation}
\label{sec:snippet-confounders}

To reduce presentation-related variation, which %
could influence comprehension~\cite{readability}, all snippets were
reformatted using Google Java Formatter~\cite{googlejavaformatter} and
displayed with a standardized IntelliJ-style theme. Participants could
choose either dark or light mode.
We did not show Javadoc comments to participants, to
avoid revealing intended functionality~\cite{identifiernames1,S4, readability}.
We also removed inline comments relevant to comprehension tasks, since tasks like code summarization and behavior
prediction can be biased by natural-language hints.
At the same time, we preserved original identifier names, including
method, parameter, and variable names. We did so to maintain
realism: in practice, developers only ever encounter code with naturally-occurring
names. %
As a result, identifier quality is part of the
overall comprehensibility signal rather than a separately controlled
factor.

\subsubsection{Background Knowledge}
\label{sec:background-knowledge}

Some snippets might require
background knowledge beyond general Java familiarity.
For example, the \emph{isValidProjectName} snippet
(\cref{fig:example_snippets}) does not require specialized knowledge,
whereas the \emph{atan2} snippet may
benefit from familiarity with trigonometry and knowledge of floating-point
behavior around singularities.
To identify such
knowledge, the authors reviewed each snippet and
enumerated domain concepts that could plausibly influence comprehension.
For each such concept, our student study included (1) a self-reported
familiarity question on a 5-point scale and (2) a multiple-choice
knowledge question shown only when a participant reported at least
some familiarity. 
We analyzed whether prior knowledge was
associated with proxy outcomes (see \cref{sec:results-background})
and found it to be mildly predictive on proxies using accuracy metrics.

\subsection{Proxy Selection and Student Study}
\label{sec:our-proxies}

It is not practical to evaluate every proxy reported in prior
work. Existing studies employ a wide variety of tasks and
measures~\cite{Wyrich:ACS23}, some of which require expensive
instrumentation (e.g., fMRI devices) or extensive interaction with
code (e.g., debugging, bug fixing, or other maintenance tasks).
We therefore focus on 14 proxies
that are both common in the literature and feasible in standard
human-subject studies without specialized equipment. Our systematic
approach for selecting these proxies is described in
\cref{sec:proxies}.

We then recruited 44 undergraduate computer science students from our two corresponding universities. These participants provided and allowed us to collect values for the
14 selected proxies on the same eight code snippets evaluated in the
expert study. \Cref{sec:student-study} provides details on the design
of this study.

\subsection{Correlation Analysis}
\label{sec:correlation}

Finally, we evaluated how well each proxy collected in the student study
aligns with the final ranking from the expert study. For each proxy, we
aggregated student responses at the snippet level by averaging proxy
values to rank the eight snippets from easiest to
hardest according to that proxy.
We then compared the rank correlation between each proxy-based ranking and the expert
ranking using \textbf{Kendall's $\tau_b$ rank
  correlation coefficient}~\cite{kendall1945treatment}. Kendall's
$\tau_b$  is well suited to our setting for three reasons. First, both
expert and proxy outputs are ordinal rankings, making a rank-based
measure appropriate. Second, with a small number of items (eight
snippets), $\tau_b$’s pairwise comparison formulation provides a
stable and interpretable measure of agreement. Third, $\tau_b$
explicitly accounts for ties in the rankings, which can arise in both
expert and proxy-derived rankings. Together, these properties
make Kendall’s $\tau_b$ an appropriate choice for quantifying how
closely each proxy reflects expert-assessed comprehension difficulty.

\section{Expert Study Design}
\label{sec:expert-study}
The goal of this study is to create an expert-determined ranking of snippets: the ``ground truth'' for evaluating the reliability of comprehension proxies.
Five professional software engineers with extensive Java experience ranked the eight code snippets from \cref{sec:snippets} based on their comprehension difficulty using a web-based application that we developed.
The study followed a multi-round structured process based on the Delphi method~\cite{delphi,delphiOriginal,randManual}.

\subsection{Participant Recruitment}
\label{sec:expert-participants}
We recruited participants through our professional networks. We contacted potential candidates via email and asked them to complete an eligibility form. %
To be included, participants had to self-report at least three years of experience in Java programming and demonstrate senior-level software engineering expertise.
We operationalized ``senior-level expertise'' as these two criteria: (i) prior promotion to a higher-level engineering role (for example, from entry level to senior roles), and (ii) experience mentoring junior engineers in a professional setting. These criteria ensure that participants not only have substantial technical experience, but have also been entrusted with increased responsibility and technical leadership within their organizations.
Candidates who reported substantial 
professional programming experience but did not report prior promotion or mentoring experience were not automatically excluded. Instead, we evaluated them based on additional background information, including job title, total years of development experience, and descriptions of the most complex systems they had worked on. Final inclusion decisions for such cases were made through discussion and consensus among the authors. Using this process, we included one participant who had substantial programming experience (10–14 years overall and 6–9 years in Java) and mentoring experience in diverse software projects, but did not explicitly report a  promotion.

We do not claim that these criteria capture all aspects of
expertise. Rather, they provide a practical way to identify
participants with substantial, relevant experience in code
comprehension.

We invited approximately 30 candidates to complete the eligibility form. Of these, 6 met the inclusion criteria, and 5 ultimately participated in the study after providing informed consent.
Four of these experts had between 10 and 20 years of professional software development experience (6 to 14 years of Java experience), three were currently working as senior software engineers and one as a non-senior engineer. The remaining expert had more than 20 years of experience in both general software development and Java, and was working as a Vice President of Engineering. Several participants have held technical leadership roles, including leading projects, mentoring engineers, and contributing to architectural decisions in large-scale systems.

The participants had experience working in organizations of varying sizes, including large-scale enterprise environments, and have contributed to software systems deployed in real-world production settings with substantial user bases. Their work includes systems that require high reliability, performance, and maintainability, such as backend services, distributed systems, and production-grade applications.
The experts primarily worked on proprietary or commercial software (5 experts), with one expert also reporting experience contributing to open-source projects.
They reported experience building diverse types of systems, including web applications (5 experts), AI based systems (4), mobile applications (2), libraries or frameworks (2), middleware (1), and desktop applications (1).
They had worked across a broad range of domains, including healthcare (3), government (3), finance or banking (2), cloud and development tools (2), as well as game development, retail or legal services, transport or airline systems, education, energy, bioinformatics, or postal systems (1 each).
We compensated each expert with a \$50 USD Amazon gift card.\looseness=-1

\subsection{Snippet Ranking Procedure}
\label{sec:expert-procedure}
We employed a three-round Delphi-style process to elicit and iteratively refine expert judgments, with the goal of obtaining a stable consensus ranking of the snippets by comprehension difficulty.
Each round was assigned a nominal one-week completion window. However, we granted extensions upon request to accommodate participants’ professional schedules and ensure thoughtful, high-quality responses. Participants were also allowed to revise and resubmit their responses before the deadline.
All interactions were anonymized to reduce potential bias arising from factors such as seniority or personality.
The stopping criteria were either: all experts converged to the same ranking; or after three completed rounds, following established recommendations that additional rounds typically yield diminishing returns~\cite{randManual}.

\subsubsection{Round 1: Independent Ranking}

In the first round, experts independently evaluated the snippets through the study website. This round consisted of three phases:

\begin{itemize}
	
	\item \textbf{Phase 1 (Independent understanding)}: Experts were presented with the eight snippets in randomized order. For each snippet, they (i) described its primary functionality and (ii) explained their assessment of its comprehension difficulty. Both tasks were completed using textual responses through the web interface. This phase ensured that experts engaged deeply with each snippet before ranking.

	\item \textbf{Phase 2 (Ranking)}: Experts ordered all snippets by comprehension difficulty using a drag-and-drop interface with eight slots.
          Participants were allowed to place multiple snippets in the same slot to indicate similar levels of comprehension difficulty.

	\item \textbf{Phase 3 (Ranking justification)}: Experts provided written justifications for their rankings, including cases where snippets were ranked as having similar difficulty.

\end{itemize}

\subsubsection{Round 2: Structured Feedback and Reranking}

After the first round, we analyzed the rankings to identify disagreements between experts. We constructed a set of \emph{disagreement points}, each capturing a specific conflict in the relative ordering of snippets, along with the differing justifications provided by experts.

In the second round, experts were presented with anonymized summaries of other participants’ assessments and the identified disagreement points. They then reviewed the rankings and justifications of other experts from the previous round. For each disagreement point, they examined the alternative perspectives and could revise their rankings and justifications. This structured feedback phase encouraged reflection and convergence while preserving independent judgment.
The reranking and revised justifications followed the same protocol and web interface as Phases 2 and 3 (ranking and justification) from the first round.

\subsubsection{Round 3: Discussion and Reranking}

The third round was conducted as a moderated, anonymized discussion using a web forum interface, as recommended by modern guidance on conducting online Delphi panels~\cite{randManual}.
We again analyzed the rankings and justifications from Round 2 to identify disagreement points. These were grouped and presented to participants in the forum. Experts first reviewed the rankings and justifications from the previous round, and then engaged in discussions with one another on each group of related disagreement points. The forum allowed experts to tag one another to send email notifications about updates.
During this process, experts could revise their rankings at any time. The moderator (the first author) facilitated the discussion by organizing disagreement points, prompting participation when needed, and encouraging experts to clarify and reflect on their reasoning, while avoiding commenting on the content of their judgments.

\subsection{Ranking Results}
\label{sec:final_rankings}

\subsubsection{Final Individual Rankings}

We encouraged but did not require the experts to reach consensus to avoid biasing their final rankings. 
Each expert's individual final ranking is in \cref{tab:final_expert_rankings}.
While the experts did not reach full consensus, their final rankings are highly aligned. Specifically, there is strong agreement on several relative positions; for example, snippets 1 and 2 are consistently ranked among the easiest, and snippets 6 and 8 among the most difficult. However, variability remains in the ordering of middle-ranked snippets and in the use of ties (by Expert A).

For the main correlation analysis, we obtained a single expert ranking by averaging ranks across experts. We tested that our results are robust under other aggregation methods (\cref{sec:aggregation_strategies}).

\begin{table}[t]
\centering
\caption{Final expert rankings of snippets after the last Delphi round.}
\small
\resizebox{\linewidth}{!}{%
\begin{tabular}{lc}
\toprule
\textbf{Expert} & \textbf{Final Ranking (Easier $\rightarrow$ More difficult to understand)} \\
\midrule
Expert A & $(1 = 2) \rightarrow 3 \rightarrow (4 = 5) \rightarrow (7 = 8) \rightarrow 6$ \\
Expert B & $1 \rightarrow 2 \rightarrow 5 \rightarrow 4 \rightarrow 7 \rightarrow 3 \rightarrow 8 \rightarrow 6$ \\
Expert C & $1 \rightarrow 2 \rightarrow 4 \rightarrow 5 \rightarrow 3 \rightarrow 7 \rightarrow 8 \rightarrow 6$ \\
Expert D & $1 \rightarrow 2 \rightarrow 5 \rightarrow 4 \rightarrow 3 \rightarrow 7 \rightarrow 8 \rightarrow 6$ \\
Expert E & $1 \rightarrow 2 \rightarrow 4 \rightarrow 5 \rightarrow 7 \rightarrow 3 \rightarrow 6 \rightarrow 8$ \\
\bottomrule
\end{tabular}%
}
\label{tab:final_expert_rankings}
\end{table}

\begin{table}[t]
	\centering
	\caption{Evolution of expert agreement across Delphi rounds. Disagreements indicate opposite pair orderings, while soft agreements capture ties versus strict ordering.}
	\small
	\resizebox{\linewidth}{!}{%
	\setlength{\tabcolsep}{5pt}
	\begin{tabular}{lcccc}
		\toprule
		\textbf{Round} & \textbf{Agreements} & \textbf{Soft Agreements} & \textbf{Disagreements} & \textbf{Total} \\
		\midrule
		Round 1 & 13 & 2 & 13 & 28 \\
		Round 2 & 17 & 1 & 10 & 28 \\
		Round 3 & 21 & 3 & 4  & 28 \\
		\bottomrule
	\end{tabular}%
	}
	\label{tab:delphi_agreement}
\end{table}

\subsubsection{Expert Agreement Across Rounds}
\label{sec:delphi_agreement}

To analyze how expert agreement evolved across rounds, we examined all pairwise comparisons between snippets (28 total pairs) and categorized them into three types based on how consistently the experts ordered them. For a snippet pair $(a, b)$, we define:
\begin{itemize}
	\item \textit{Agreement}: all experts assign the same relation between $a$ and $b$, either $a < b$, $b < a$, or $a = b$.
	
	\item \textit{Soft agreement}: there is no direct conflict in ordering, but not all experts assign the same relation. In particular, some experts assign a strict ordering (either $a < b$ or $b < a$, but all experts agree on the same direction) while others assign a tie ($a = b$).\looseness=-1 %
	
	\item \textit{Disagreement}: there is a direct conflict in ordering, with some experts assigning $a < b$ and others assigning $b < a$.
\end{itemize}
\Cref{tab:delphi_agreement} shows how experts converged towards more consistent, strict orderings in later rounds.
Inter-expert agreement, measured using Kendall’s coefficient of concordance ($W$)~\cite{kendall1939distribution} computed using the corresponding Friedman test formulation~\cite{friedman1937use}, shows a steady increase across rounds: $W = 0.7358$ in Round 1, $W = 0.8472$ in Round 2, and $W = 0.9338$ in the final round. These results indicate that expert rankings became increasingly concordant after each round, as expected from the Delphi method.

Overall, the results suggest that the Delphi process reduces conflicts and leads to a largely consistent ordering with high agreement, while retaining limited areas of ambiguity rather than enforcing complete consensus across all snippets.

\section{Student Study Design}
\label{sec:student-study}

The goal of this study is to collect well-established proxies like those in prior comprehension studies, on the same set of code snippets used in the expert study. %
The study uses a within-subjects design: each participant interacted with each code snippet. For each snippet, participants completed a set of comprehension tasks, with measures including response time, answer correctness, and subjective ratings.

\subsection{Participant Recruitment}

Participant recruitment targeted undergraduate Computer Science (CS) students from our two universities. 
Using flyers, email announcements, and in-class invitations, we recruited 44 students: 37 from one institution and 7 from the other. 
Students were compensated with food after the study, and an entry into a raffle for two \$50 USD Amazon gift cards. %
Participants were required to be at least 18 years old, have prior experience programming in Java, and have completed or be enrolled in advanced programming courses. %
The participant pool consisted of 33 male, 10 female, and one non-binary individual. On average, participants reported 3.1 years of Java programming experience. %
There were 16 4th-year students, 15 third-years, 11 second-years, and 2 first-years.

\subsection{Experimental Procedure}

Participants completed the study individually using their own laptops via a dedicated study website we developed, in-person in a controlled lab environment. %
Sessions lasted at most 90 minutes, consistent with prior comprehensibility research on sustained attention~\cite{52,33-13,33-24},
though most finished within 70 minutes.

Participants were first verbally introduced to the study and gave informed consent.
They then completed a short demographic and programming background questionnaire, which asked about their perceived programming experience relative to their classmates,
their experience with Java and object-oriented programming generally, and their overall programming experience; we analyze whether these factors impact the main results
in \cref{sec:results-background}.
Participants then completed a brief practice task to familiarize them with the web interface and question format and reduce learning effects unrelated to code comprehension.

Participants then completed the main study. The eight Java code snippets were presented in a different random order to each participant.
For each snippet, participants first read the code and then answered four comprehension questions presented sequentially in randomized order.
Once a participant advanced, they could not return to previous questions or snippets, preventing answer revision and reducing potential carryover effects across tasks.
The code remained visible while answering the questions to avoid introducing memory-related confounds into our results.

Participants were instructed not to use external resources (e.g., search engines or AI tools) so that the collected measures reflect their own comprehension effort.
An on-screen timer timed all interactions. %
Participants were instructed to pause the timer whenever they were not actively working on the tasks, and to resume it when continuing.
Researchers monitored sessions in-person to ensure compliance with these instructions and to assist with technical or logistical issues. %
To reduce fatigue, the study system enforced short breaks after every two snippets, during which participants could resume the study after 30 to 60 seconds.
A longer break was also enforced at the midpoint of the session, with a duration ranging from one to seven minutes. 

After completing the comprehension tasks, participants completed a post-study survey assessing their familiarity with the background concepts required for each snippet (\cref{sec:background-knowledge}). %

\subsection{Comprehension Proxies}
\label{sec:proxies}

We wanted a broad set of common comprehension proxies from the literature that are feasible for students in a single study session.
Recall that a comprehension proxy is a pair of a specific task (\eg writing a summary) and a measure (\eg time or correctness).

\subsubsection{Comprehension Tasks}

The taxonomy proposed by Wyrich \etal~\cite{Wyrich:ACS23} organizes tasks into four categories: providing information about code (Tc1), providing subjective opinions (Tc2), debugging (Tc3), and maintenance (Tc4) (see \cref{tab:proxies-wyrich}). As discussed in \cref{sec:rw-tasks}, prior work is heavily concentrated on Tc1 and Tc2 tasks, so we focus on those and exclude Tc3 and Tc4 tasks.
Tc1 and Tc2 tasks are widely-used and low-cost to administer, unlike Tc3 and Tc4 tasks.
Tc1 tasks are much more common than Tc2 tasks in the literature, so we include several Tc1 tasks and one Tc2 task.
We selected representative tasks within Tc1 via our open-coding analysis of comprehension questions from prior studies (see \cref{sec:open-coding}), which organizes questions into categories such as program function, semantics, and syntax. To ensure conceptual coverage while keeping the study feasible, we selected one representative task from each of these categories. 
Specifically, we included these tasks:
\begin{itemize}
    \item \textbf{Tc1: Program function}: describe what the code does. %
    \item \textbf{Tc1: Semantics}: determine the code's output for a given input. %
    \item \textbf{Tc1: Syntax}: identify syntactic elements in the code. %
    \item \textbf{Tc2: Subjective rating}: rate comprehension and task difficulty. %
\end{itemize}
\paragraph{Question Design}
\label{sec:question-design}
For the \textbf{Program function} and \textbf{Subjective rating} tasks, all snippets share a fixed question format.
For the \textit{program function} task, participants described the overall functionality of the method in their own words. %
The \textit{subjective task} used two 5-point Likert scales to assess perceived comprehension and task difficulty, ranging from ``Very difficult'' to ``Very easy.''

For the \textbf{Semantics} and \textbf{Syntax} tasks, we used a semi-structured design process informed by our open-coding analysis of prior studies (\cref{sec:open-coding})
to design unique questions for each snippet.
We first derived question templates from %
the literature:
for the \textit{semantics task}, the template required computing the return value for a specific input;
for the \textit{syntax task}, the template required identifying or locating a syntactic element within a local code region (e.g., conditions, variable usage, or control structures).
We then used ChatGPT to fill in these templates to generate candidate questions for each snippet. %
All generated questions were manually reviewed by the research team to ensure correctness, clarity, and consistency across snippets. 
We excluded questions that were ambiguous, misleading, or that revealed the intended functionality of the code. 
From the remaining candidates, we randomly selected one question per task category for each snippet.
\looseness=-1
Ground-truth answers for all questions were established by the research team and verified through manual inspection of the code to support correctness-based measures used in our analysis.

As a concrete example, for the \emph{isValidProjectName} snippet
(\cref{fig:example_snippets}), participants were asked: (\emph{output})
\textit{``What value does the method return when called with input
\texttt{"a1\_"}, given \texttt{MAX\_NAME\_LENGTH = 10} and
\texttt{VALID\_NAME\_SET = \{'\_', '-', '.'\}}?''}; (\emph{syntaxBL})
\textit{``Is \texttt{MAX\_NAME\_LENGTH} used in a conditional
expression?''}; (\emph{function}) \textit{``What does this code snippet
do? Briefly describe its main functionality.''}; and (\emph{scaleSM},
\emph{scaleST}) \textit{``How easy or difficult was this snippet to
understand?''} and \textit{``How easy or difficult was this task?''}
(Very easy -- Very difficult). The full set of tasks and questions for all eight
snippets is available in our replication package~\cite{replication-package}.

\begin{table}[t]
\scriptsize
\centering
\caption{Comprehensibility proxies used in our study. Direction indicates whether higher values correspond to harder or easier comprehension difficulty. sec = seconds.}
\label{tab:proxies}

\begin{tabularx}{\columnwidth}{l X l l}
\toprule
\textbf{Proxy} & \textbf{Description} & \textbf{Scale} & \textbf{Direction} \\
\midrule

\multicolumn{4}{l}{\textbf{Time-based}} \\

readTime & Reading time before answering questions & sec & $\uparrow$ harder \\

time\_function & Time to write function summary & sec & $\uparrow$ harder \\

time\_output & Time to predict output given an input & sec & $\uparrow$ harder \\

time\_syntaxBL & Time to answer syntax question & sec & $\uparrow$ harder \\

time\_correct\_function & Time to write summary graded 4 or 5 & sec & $\uparrow$ harder \\

time\_correct\_output & Time for correct output predictions & sec & $\uparrow$ harder \\

time\_correct\_syntaxBL & Time for correct answers to syntax question & sec & $\uparrow$ harder \\

time\_scaleSM & Time to decide snippet difficulty rating & sec & $\uparrow$ harder \\

time\_scaleST & Time to decide task difficulty rating & sec & $\uparrow$ harder \\

\midrule

\multicolumn{4}{l}{\textbf{Correctness-based}} \\

acc\_function & Function summary grade & 1--5 & $\uparrow$ easier \\

acc\_output & Output prediction correctness & 0/1 & $\uparrow$ easier \\

acc\_syntaxBL & Syntax question correctness & 0/1 & $\uparrow$ easier \\

\midrule

\multicolumn{4}{l}{\textbf{Rating-based}} \\

scaleSM & Snippet difficulty rating & 1--5 & $\uparrow$ harder \\

scaleST & Task difficulty rating & 1--5 & $\uparrow$ harder \\

\bottomrule
\end{tabularx}
\end{table}

\subsubsection{Comprehension Measures}
\label{sec:our-measures}

We used three types of measures:
\begin{itemize}
    \item \textbf{Time}: reading time per snippet and time spent on each task.
    \item \textbf{Correctness}: correctness of responses to Tc1 questions.
    \item \textbf{Subjective ratings}: answers to Tc2 Likert-scale questions.
\end{itemize}
These are the three most common measures in the literature (\cref{sec:rw-measures}).
We excluded physiological measures as they require
specialized equipment and expertise, and are less common.
We also included combined measures based on the time for \emph{correct} answers to questions, which have been used in prior work~\cite{33-9, von1995program}.

Correctness was determined differently depending on the question type. 
For the semantics and syntax questions, which were binary, we defined correctness based on predefined ground-truth answers validated by the research team. 
For program function questions, we used a 5-point ordinal scale (adapted from prior work~\cite{S87}) to assess the quality of free-response answers.

Grades of 4 or 5 were considered ``correct'' for combined measures.
We used each method's Javadoc to create the initial grading rubric.
To ensure grading consistency, %
all authors first independently graded a shared sample of responses and then discussed discrepancies to finalize the rubric.
Subsequently, each response was graded independently by two authors, and disagreements were resolved through discussion. All grades, justifications, and discussion logs are in our artifact~\cite{replication-package}.
Inter-rater agreement for program function grading, based on the 5-point ordinal scale for free-response function summaries, was Cohen's $\kappa = 0.57$.

\subsubsection{Summary of Proxies}

The final set of proxies includes 14 combinations of the selected task types (program function, semantics, syntax, subjective rating) and measures (time, correctness, and derived measures). For time-based and subjective-rating proxies, higher values indicate greater comprehension difficulty, whereas for correctness-based proxies, lower values indicate greater difficulty. \Cref{tab:proxies} lists all proxies used in our study.

\section{Results}
\label{sec:evaluation}
\label{sec:results}

\begin{figure}[t]
	\centering
	\includegraphics[width=0.8\linewidth]{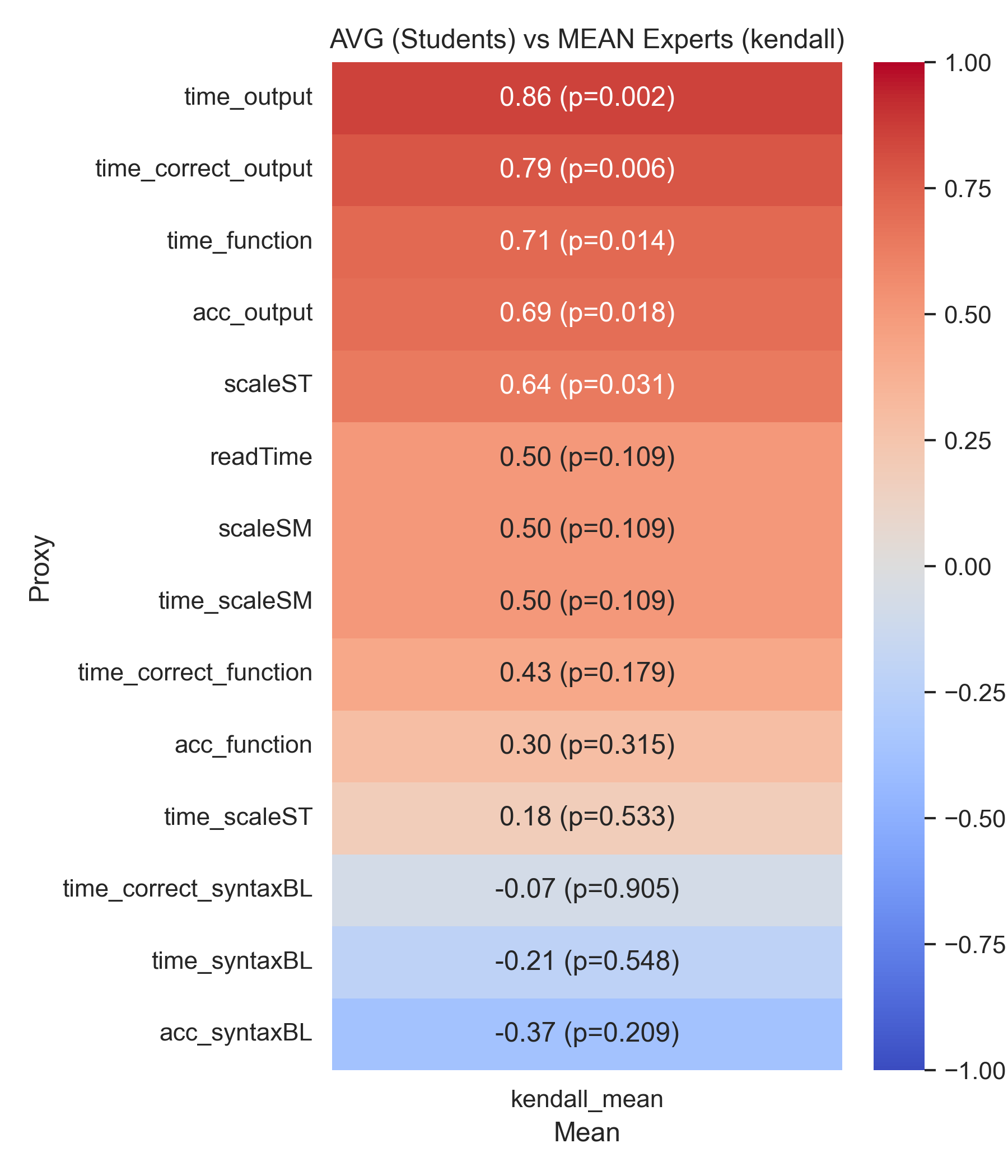}
	\caption{Correlation between proxies and expert-determined rankings, ordered by strength of correlation (strongest at the top). Descriptions of
		the proxies are in \cref{tab:proxies,sec:proxies}.}
	\label{fig:heatmap_majority_kendall}
\end{figure}

Figure~\ref{fig:heatmap_majority_kendall} presents the results of the correlation analysis between proxies and the expert-determined ranking. 
Time-based semantic proxies exhibit strong alignment with expert-defined comprehensibility, with \emph{time\_output} achieving a correlation of 0.86, followed by \emph{time\_correct\_output} (0.79) and \emph{time\_function} (0.71). These values indicate that these proxies approximate expert judgment well.
The five top proxies are statistically significant at
$\alpha = 0.05$ (e.g., \emph{time\_output}, $p=0.002$). The remaining
proxies do not reach statistical significance at this sample size, but
their relative ordering remains stable across the secondary analyses
in \cref{sec:secondary-analyses} (alternative aggregation strategies and
individual-level data), increasing confidence in the overall
proxy-reliability pattern.

In contrast, syntax-based proxies fail to capture expert judgments. Measures such as \emph{acc\_syntaxBL} ($-0.37$) and \emph{time\_syntaxBL} ($-0.21$) show weak or even negative correlations, suggesting that the syntax tasks reflect a different construct than expert-perceived comprehensibility.
More interestingly, proxies in the middle show moderate but consistent alignment. Perception-based measures such as \emph{scaleST} (0.64) and \emph{scaleSM} (0.50), as well as general effort measures such as \emph{readTime} (0.50), exhibit meaningful correlations with expert rankings, though substantially weaker than semantic time-based proxies. These results suggest that while subjective ratings and general reading effort capture some aspects of perceived difficulty, they do not reflect the same depth of understanding as time spent solving semantic tasks. Similarly, time spent on task difficulty judgments (e.g., \emph{time\_scaleST}, 0.18) shows only weak alignment, indicating that deliberation about perceived difficulty is a less reliable signal.
Overall, these results demonstrate a clear gradient: semantic time-based proxies provide the strongest approximation of expert-defined comprehensibility, perception-based and general effort measures provide moderate signals, and syntax-based measures correlate poorly.
\looseness=-1

\subsection{Secondary Analyses}
\label{sec:secondary-analyses}

This section details secondary analyses that we performed to ensure that the result is stable, especially with respect to possible confounders and degrees of freedom in our experimental design. More details on all of these analyses are in Appendix A in our supplementary material.

\subsubsection{Aggregation Strategies}
\label{sec:aggregation_strategies}

Our experts did not reach complete agreement, so we aggregated their rankings into a single reference ordering by assigning each snippet its average rank across experts. %
To ensure that the results were stable, we also considered alternative aggregation strategies: requiring full agreement, allowing ties without contradiction, and permitting a small number of inconsistencies. The overall pattern of which proxies are well- or poorly-correlated with the expert-derived ranking remains stable across all aggregation strategies.
We also confirmed that using individual experts’ rankings instead of an aggregated ranking leads to consistent results: the relative performance and ordering of proxies remain largely unchanged across experts.

Our correlations are also based on aggregated student data. We tested whether analyses using individual student data yield similar patterns of correlations.
They do: the relative ordering of proxies
remains consistent. Per-student correlations do exhibit greater variability and are lower, due to individual differences %
and noise.

\subsubsection{Student Characteristics and Fatigue}
\label{sec:results-background}

Differences in participant background or task order could explain variation in the comprehension proxies, and therefore potentially confound our main results. To assess possible effects, we fit regression models relating participant characteristics and task position to each proxy. %
Overall, participant characteristics exhibit limited and task-specific associations with comprehension performance. There are a few statistically significant relationships for accuracy on output and program function tasks, particularly for \emph{Programming Experience Relative to Peers} ($\beta \approx 0.33$, $p = 0.024$ for program function tasks; $\beta \approx 0.36$, $p = 0.045$ for output tasks), \emph{Object-Oriented Experience} ($\beta \approx 0.39$, $p = 0.026$ for program function tasks; $\beta \approx 0.45$, $p = 0.023$ for output tasks), and \emph{Background Knowledge} (for program function tasks, $\beta \approx 0.35$, $p = 0.010$). In contrast, there are no consistent effects for syntax tasks or for most general experience measures like \emph{Java Experience}.\looseness=-1

Time-based proxies show little systematic relationship with participant characteristics, with effect sizes generally small and non-significant across tasks. However, \emph{Fatigue} (operationalized as task order) has a strong and consistent effect on timing: participants complete tasks more quickly as the study progresses (e.g., $\beta \approx -7.9$ for \emph{time\_function}, $p < 10^{-3}$; $\beta \approx -15.6$ for \emph{snippet\_total\_time}, $p < 10^{-8}$). This effect is not accompanied by a corresponding decrease in accuracy (all $p > 0.05$).

Taken together, these results suggest that while certain background factors are weakly associated with performance on specific comprehension tasks, overall comprehension outcomes are not strongly driven by participant characteristics. The observed changes in timing over the course of the study are more consistent with adaptation or increasing familiarity with the task format than with fatigue-related degradation in performance.

\section{Discussion and Implications}
\label{sec:discussion}

\subsection{Implications for Interpreting Prior Work}
\label{sec:implications}

Our findings on proxy reliability have important implications for how prior work on
program comprehension should be interpreted.

\textbf{Syntax-only proxies.} A subset of prior studies relies %
on syntax tasks like identifying structural properties or locating
code elements, typically with response time or accuracy
measures. %
Although some studies frame the construct as
\emph{readability}, they operationalize it the same way as studies of program
comprehension \cite{Wyrich:ACS23}.
For example, Asenov et al.~\cite{33-29} aim to evaluate whether richer code
visualizations improve comprehension, measuring how quickly and
accurately developers answer yes/no questions about code structure;
similarly, Tenny~\cite{33-4} investigates readability by assessing how accurately
participants answer questions about program properties after reading
the code.
These task–measure combinations are syntax
proxies (\emph{time\_syntax} and \emph{acc\_syntax}) that our results show
are unreliable signals of comprehensibility.
The improvements these studies report may reflect increased efficiency in detecting or extracting
structural features and not genuine improvements in
understanding program behavior, and so should be interpreted
with caution.
\looseness=-1

\textbf{Aggregated syntax-semantic proxies.} Another group of studies~\cite{33-15,penningtonModel,S75,S4} uses a mix of syntax and semantics tasks,
but aggregates performance across them into
one correctness or time measure. %
These studies include reliable tasks that ask about program behavior, but the
aggregation implicitly treats all task–measure combinations as
equally valid proxies of comprehensibility. Our findings indicate
that this assumption does not hold, since proxies vary a lot in
their alignment with expert-determined comprehensibility:
syntax proxies provide weaker signals than semantic ones.
Aggregating across heterogeneous proxies can obscure meaningful differences, and
so reported effects may reflect limitations of the
measurement approach rather than the absence of an effect on
comprehensibility itself.

\textbf{Accuracy-only proxies.} Some studies~\cite{S24, S35, S38} employ tasks that directly target program
understanding, such as explaining the purpose of code, but rely
primarily on correctness as the sole measure. These task–measure
combinations are proxies similar to \emph{acc\_function}.
While these tasks are well aligned with the construct of
comprehensibility, our findings show that accuracy alone provides
a weaker and less consistent signal compared to a time-based measure
of the same task (\ie \emph{time\_function}).
As a result, these studies may
detect whether participants reach a correct interpretation, but
may underestimate differences in how difficult the code is to
understand.

\textbf{Semantic proxies.} In contrast, some studies \cite{S39, S48, S50, S57, S59, S60} 
use tasks that require reasoning
about program behavior, such as predicting outputs or determining
intermediate values during execution, measured via both response
time and accuracy.
These task–measure combinations align closely with proxies that
show strong and consistent agreement with expert-determined
comprehensibility in our evaluation (e.g., \emph{time\_output}).
This alignment indicates that such proxies more effectively capture
the cognitive effort required to understand program behavior.
Therefore, findings from these studies are more likely to reflect
genuine differences in comprehensibility rather than artifacts of
the measurement approach.
\looseness=-1

Overall, these observations highlight that the validity of conclusions
in program comprehension research depends on the
alignment between the chosen task–measure combinations and the
underlying construct of comprehensibility.
Reliable and meaningful results require careful selection and interpretation of proxies.

\subsection{Implications for Future Studies}
\label{sec:future-work}

Our results suggest that future studies can approximate code
comprehensibility using simple and cost-effective proxies
based on response time for input-output
reasoning tasks. These proxies show strong and consistent
alignment with expert judgments, indicating that they capture
key aspects of the cognitive effort required to understand
program behavior.
Importantly, when using such reliable proxies, studies with
student participants meaningfully approximate
professional engineers’ assessments of comprehensibility. So,
researchers can continue to use students as a practical and scalable
study population, as long as appropriate task–measure
combinations are selected---that is, (expensive) expert participants are not necessary for reliable results.
But, our findings caution
against relying on proxies that focus on syntax or use accuracy
alone, as these provide weaker, less consistent signals of the
underlying construct.

\subsection{Limitations and Threats to Validity}
\label{sec:limitations}

\textbf{Construct validity.}
Our study operationalizes code comprehensibility using
expert-derived rankings as a proxy for ground truth. While
expert agreement provides a practical and principled
approximation, it is not perfect.
Different expert populations or elicitation procedures
may yield different rankings, though the Delphi protocol mitigates the latter.
Our selected proxies may miss aspects of comprehension, such as long-term retention or
maintainability. Correctness for function summaries is based on human grading on
a 5-point scale, which introduces subjectivity despite calibration and
moderate inter-rater agreement ($\kappa = 0.57$).
Grading disagreements were resolved through a rigorous protocol; however, only
\emph{acc\_function} and \emph{time\_correct\_function} depend on these grades,
and neither was among the top-performing proxies.
Our taxonomy of comprehension tasks is based on an open-coding
process over prior studies. Although we followed a systematic
procedure and achieved high inter-rater agreement ($\kappa =
0.89$), the resulting categorization may be subjective
and may not capture every task in the literature.

\textbf{Internal validity.}
Several factors may influence the observed proxy values
independent of comprehension. For example, participant fatigue,
learning effects, or familiarity with the study interface may
affect response times, although our analysis in \cref{sec:results-background} suggests that time
decreases are more consistent with adaptation than degradation in
performance. Noise may also arise from self-monitored
timing (\eg pausing the timer).

\textbf{External validity.}
Our study uses just eight Java methods selected under
controlled criteria (e.g., size, use of standard libraries,
absence of project-specific context). To reduce required
background knowledge, we only included methods that exclusively use standard Java types.
To reduce dependence on external project context and object state, and
consequently enable more controlled comparisons across snippets, we focused
on static methods.
While these choices improve experimental control,
they exclude methods that depend on custom
types, frameworks, or richer context, and thus may limit
generalizability. %
Similarly, although our expert panel consists of experienced professional
developers, the number of experts is limited and may not fully represent
the diversity of industry practices. Nevertheless, we verified
that, when correlating student-derived proxies with individual
expert rankings (rather than aggregated rankings), the overall
trends remain consistent. Finally, we did not evaluate all
possible comprehension proxies, but instead focused on a
representative subset of frequently used task–measure combinations
from prior work. %

\section{Conclusion}

We investigated the reliability of comprehension
proxies from the literature with two complementary studies:
a Delphi study of expert software engineers to establish
ground truth, and a study of students with common proxies on the same code snippets.
We correlated the results of the two studies, and showed
that time measures and tasks that require semantic reasoning
are the most reliable, while syntax tasks are unreliable.
Future code comprehension studies should use our results
to guide their choices of proxies.

\begin{acks}
  We thank the experts and students who participated in our study. This work was supported in part by National Science Foundation grants CCF-2239107, CCF-2414110, and CCF-2414111.
  The views expressed herein are the authors' own and do not necessarily reflect those of our funders.
\end{acks}

\section*{Data Availability Statement}
Our replication package is publicly available~\cite{replication-package}.

\balance
\bibliographystyle{ACM-Reference-Format}
%
%

%

\newpage
\appendix
\section{More Details on Secondary Analyses}
\label{app:secondary-analyses}

\subsection{Aggregation Strategies for Expert Rankings}
\label{app:aggregation_strategies}

To account for residual variability across expert judgments, we construct multiple partial rankings using different aggregation strategies, as summarized in Table~\ref{tab:expert_aggregation}. These strategies capture varying levels of consensus, ranging from strict agreement to more relaxed formulations that allow ties or limited violations.

The number of resulting rankings differs substantially across strategies, reflecting the degree of flexibility each approach permits. Together, these aggregated rankings provide a principled way to model expert uncertainty and are used as alternative ground truths in our subsequent analysis.

\begin{table}[t]
\centering
\scriptsize
\begin{tabular}{lcl}
\toprule
\textbf{Strategy} & \textbf{\# Rankings} & \textbf{Representative Rankings} \\
\midrule
Strict & 4 &
$\{[1,4,7,6], [1,5,7,6], [2,4,7,6], [2,5,7,6]\}$ \\

Unanimous up to ties & 6 &
$\{[1,2,4,7,6], [1,2,4,7,8], [1,2,5,3,6], \dots\}$ \\

1 violation & 14 &
$\{[1,2,4,5,7,6], [1,2,4,5,7,8], [1,2,4,7,6,8], \dots\}$ \\

Majority/mean & 1 &
$\{[1,2,5,4,3,7,8,6]\}$ \\
\bottomrule
\end{tabular}
\caption{Partial rankings derived from different expert aggregation strategies. The table reports the number of valid rankings per strategy and representative examples; full sets are used in the analysis.}
\label{tab:expert_aggregation}
\end{table}

\subsubsection{Expert Ranking Aggregation and Correlation Analysis}

\paragraph{Expert Rankings}
Let
\[
S = \{1, 2, \dots, 8\}
\]
denote the set of code snippets used in the study.

Each expert provides a ranking over the snippets, allowing for ties.  
We model an expert’s judgment as a binary preference relation
\[
\prec_e \;\subseteq\; S \times S,
\]
where \( a \prec_e b \) means that expert \( e \) considers snippet \( a \) easier to understand than snippet \( b \).
If an expert judges two snippets to be equally difficult, neither ordering is included.

Given a candidate aggregated ranking
\[
R = (r_1, r_2, \dots, r_k),
\]
we say that \( R \) induces the pairwise relation
\[
r_i \prec_R r_j \quad \text{iff} \quad i < j.
\]

Our goal is to construct aggregated expert rankings under different consistency assumptions and then correlate them with rankings derived from student data. In our setting, the rankings produced by Strategy~4 (pure majority aggregation) and Strategy~5 (mean-rank aggregation) are identical. Therefore, we report their results jointly under the label \emph{majority/mean}.

\subsubsection{Aggregation Strategies for Expert Rankings}

We consider five aggregation strategies, each imposing different constraints on how expert preferences must align with the aggregated ranking.

\paragraph{Strategy 1: Strict Unanimous Agreement}
A ranking \( R \) satisfies \emph{strict unanimous agreement} if for every induced pair \( (a,b) \in \prec_R \),
\[
\forall e \in E:\quad a \prec_e b.
\]

That is, all experts must strictly agree on the ordering of every pair in \( R \).
Ties are not permitted for any included comparison.
This is the strongest notion of consensus and typically yields short rankings.

\paragraph{Strategy 2: Unanimous Agreement up to Ties}
A ranking \( R \) satisfies \emph{unanimous agreement up to ties} if for every induced pair \( (a,b) \in \prec_R \),
\[
\forall e \in E:\quad \neg(b \prec_e a).
\]

Experts may either agree that \( a \) is easier than \( b \) or consider them tied, but no expert may express the opposite preference.
This strategy relaxes strategy~1 by allowing ties while still excluding contradictions.

\paragraph{Strategy 3: Unanimous up to Ties with One Global Violation}
A ranking \( R \) satisfies strategy~3 if it violates the strategy~2 condition for \emph{at most one unordered pair} \( \{a,b\} \subseteq R \).
Formally, there exists at most one pair such that
\[
\exists e \in E:\quad b \prec_e a,
\]
while all other pairs satisfy the strategy~2 constraint.

This strategy captures rankings that are nearly unanimous while permitting a single global inconsistency, enabling longer rankings.

\paragraph{Strategy 4: Pure Majority Aggregation}
For any pair \( (a,b) \), define the majority preference relation:
\[
a \prec_{\text{maj}} b
\quad \text{iff} \quad
|\{e \mid a \prec_e b\}| > |\{e \mid b \prec_e a\}|.
\]

A ranking \( R \) is valid under pure majority aggregation if all its induced pairs are consistent with \( \prec_{\text{maj}} \).
This approach allows disagreement among experts and corresponds to classical majority voting.
Because majority relations may contain cycles, not all snippets can necessarily be included.

\paragraph{Strategy 5: Mean-Rank (Borda-Style) Aggregation}
Let \( \text{rank}_e(a) \) be the numeric rank assigned to snippet \( a \) by expert \( e \), with tied snippets receiving equal ranks.
The mean rank of snippet \( a \) is
\[
\overline{\text{rank}}(a) = \frac{1}{|E|} \sum_{e \in E} \text{rank}_e(a).
\]

The aggregated ranking is obtained by sorting snippets in increasing order of \( \overline{\text{rank}}(a) \).
This method produces a single total order over all snippets and serves as a baseline aggregation strategy.

\subsubsection{Selection of Expert Rankings for Analysis}

For each aggregation strategy, we enumerated all valid rankings and selected those with \emph{maximum length} (i.e., containing the largest number of snippets).
These longest rankings were used for correlation analysis, as they maximize comparability with student-derived rankings while respecting the defining constraints of each strategy.

\subsubsection{Results}

\begin{figure}[t]
    \centering
    \includegraphics[width=0.9\linewidth]{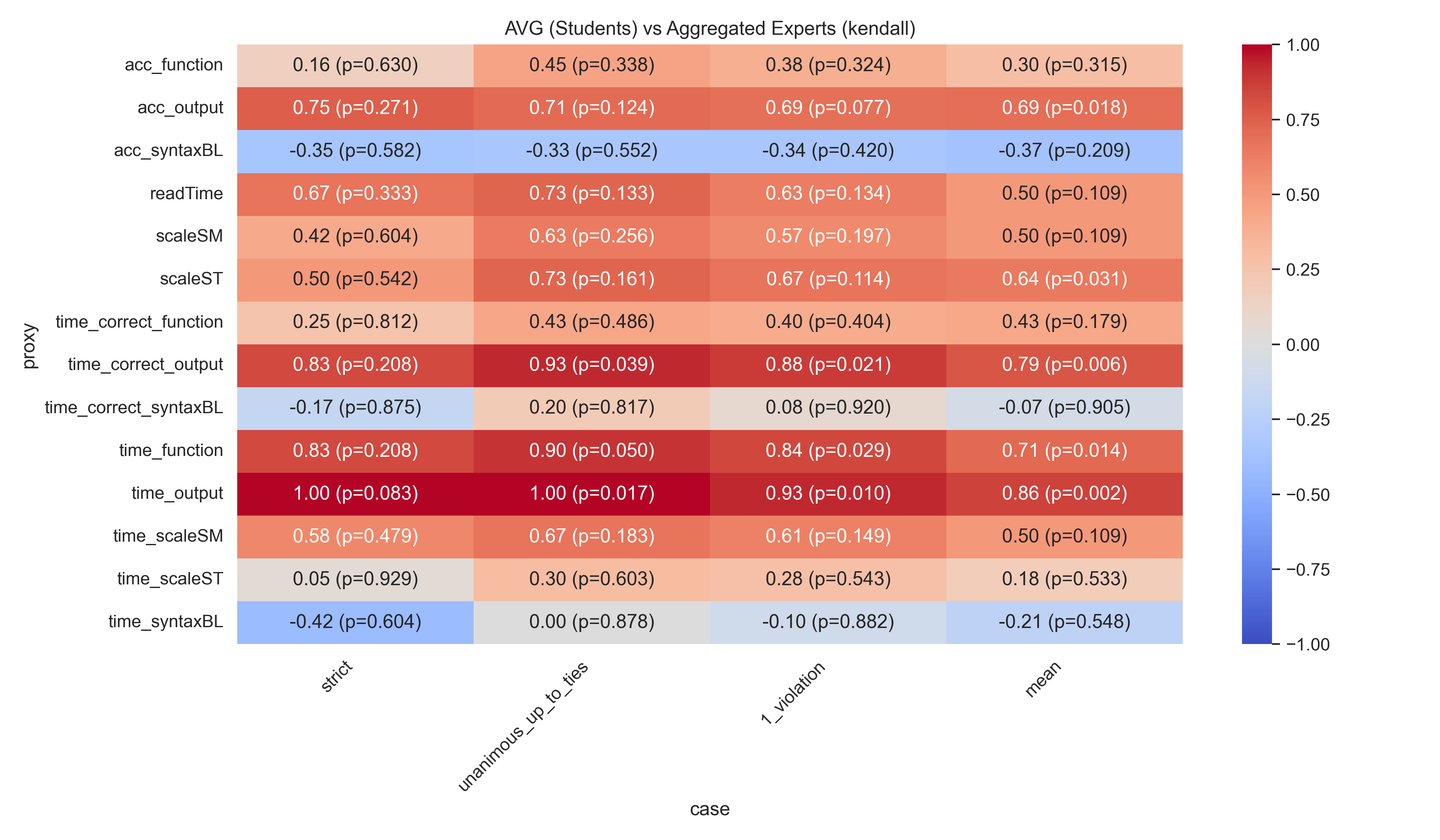}
    \caption{Correlation between aggregated student proxies and expert rankings across different expert aggregation strategies (Kendall).}
    \label{fig:heatmap_agg_vs_agg_kendall}
\end{figure}

\begin{figure}[t]
    \centering
    \includegraphics[width=0.9\linewidth]{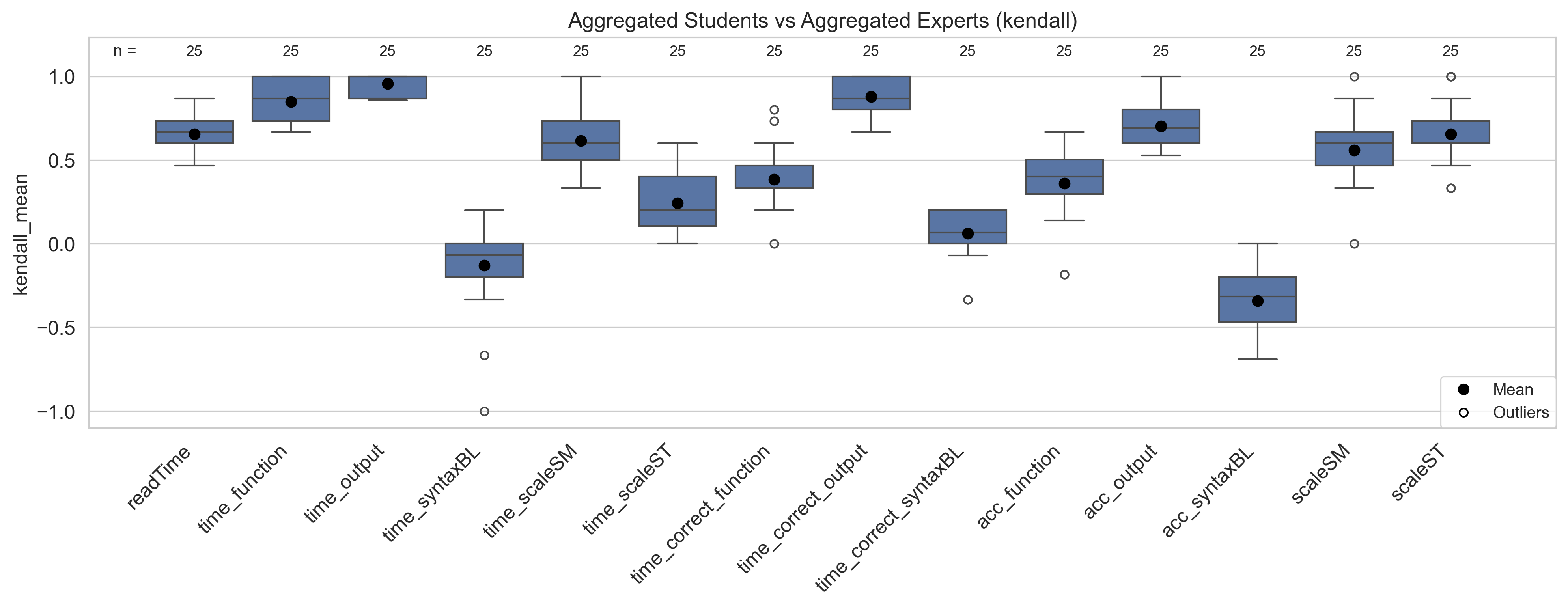}
    \caption{Distribution of correlations between aggregated student proxies and expert rankings across aggregation strategies (Kendall).}
    \label{fig:box_agg_vs_agg_kendall}
\end{figure}

Figures~\ref{fig:heatmap_agg_vs_agg_kendall} and~\ref{fig:box_agg_vs_agg_kendall} show results across different expert aggregation strategies. While absolute correlation values vary, the relative ranking of proxies remains stable across aggregation methods.

\subsection{No Aggregation in Expert Side}
\label{app:no-aggregation}

\begin{figure}[t]
    \centering
    \includegraphics[width=0.9\linewidth]{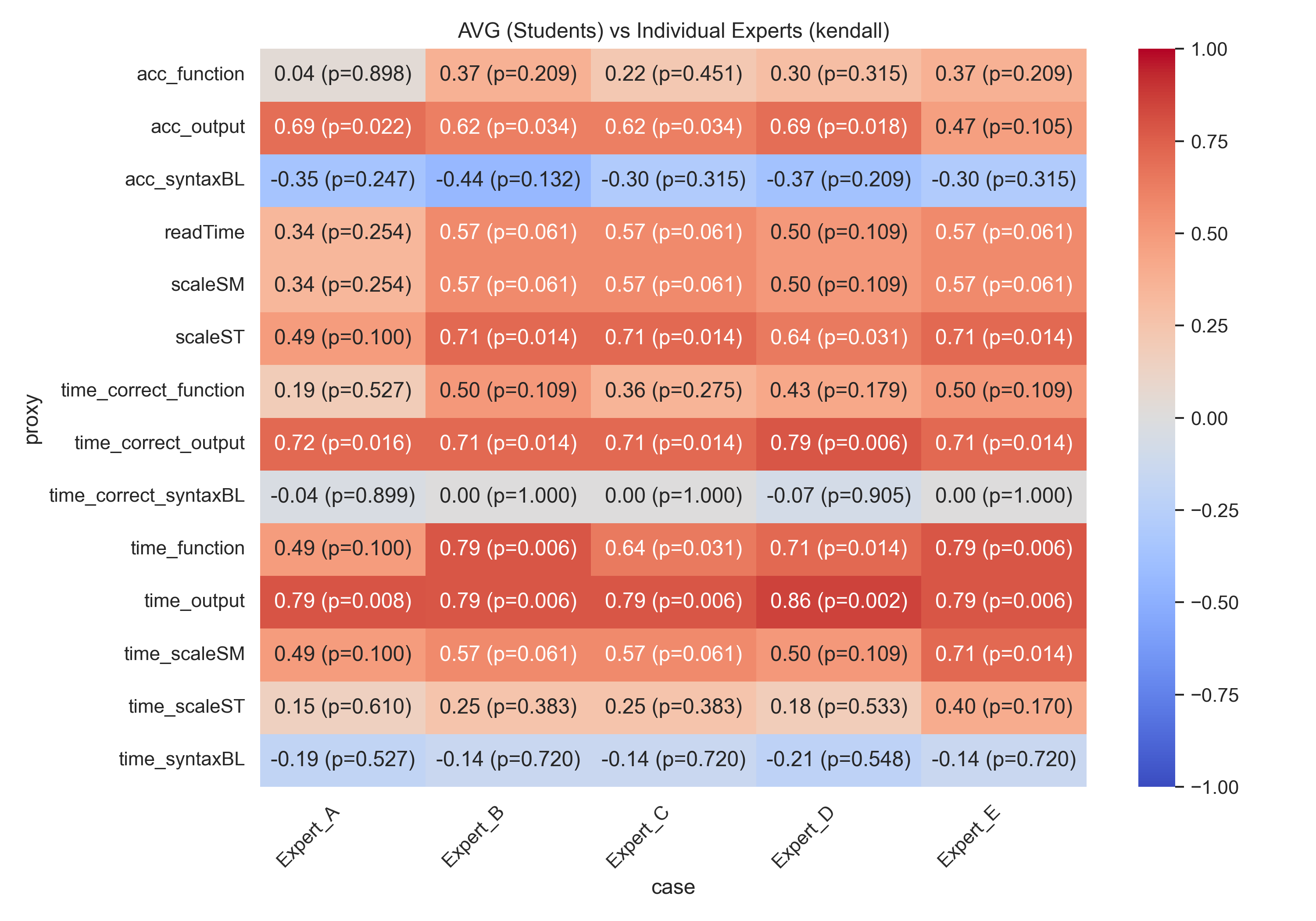}
    \caption{Correlation between aggregated student proxies and individual expert rankings (Kendall).}
    \label{fig:heatmap_avg_vs_expert_kendall}
\end{figure}

\begin{figure}[t]
    \centering
    \includegraphics[width=0.9\linewidth]{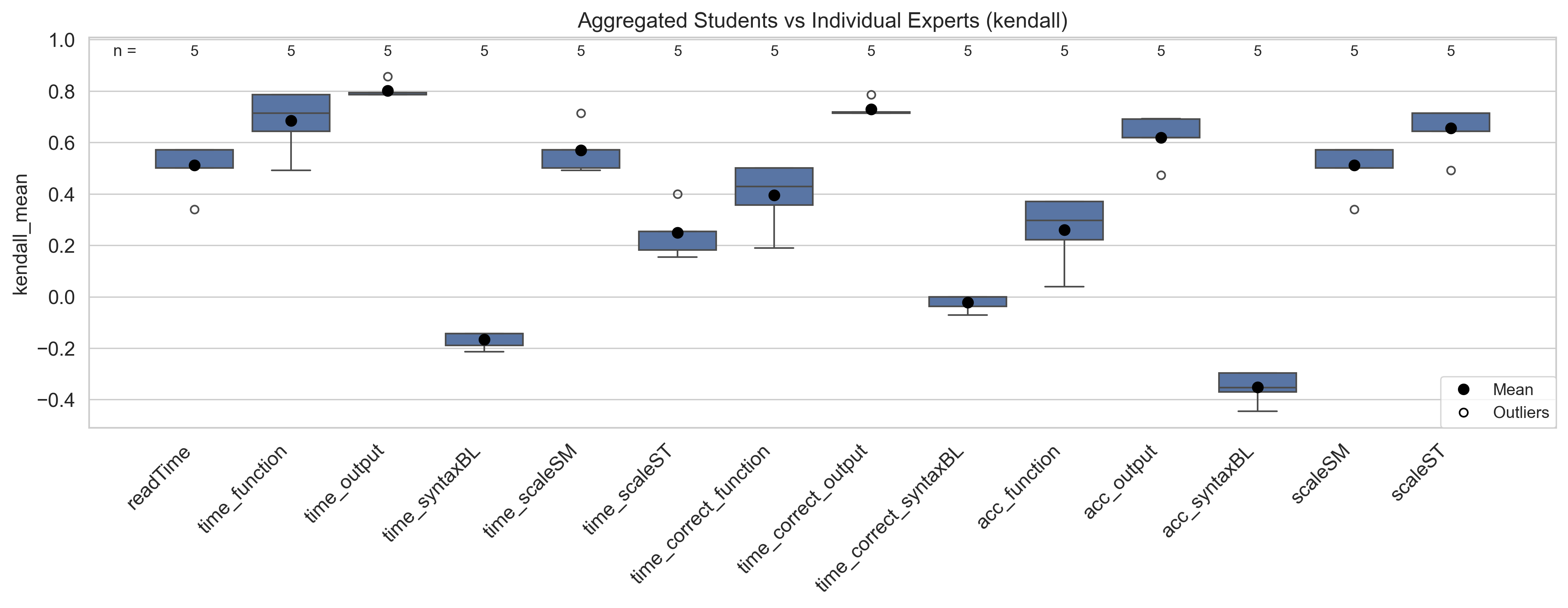}
    \caption{Distribution of correlations between aggregated student proxies and individual expert rankings (Kendall).}
    \label{fig:box_avg_vs_expert_kendall}
\end{figure}

Figures~\ref{fig:heatmap_avg_vs_expert_kendall} and~\ref{fig:box_avg_vs_expert_kendall} compare aggregated student proxies against individual experts. The overall trends remain consistent, indicating that the results are not driven by a specific expert.

\subsection{No Aggregation in Student Side (vs Aggregated Experts)}
\label{app:student-no-agg}

\begin{figure}[t]
    \centering
    \includegraphics[width=0.9\linewidth]{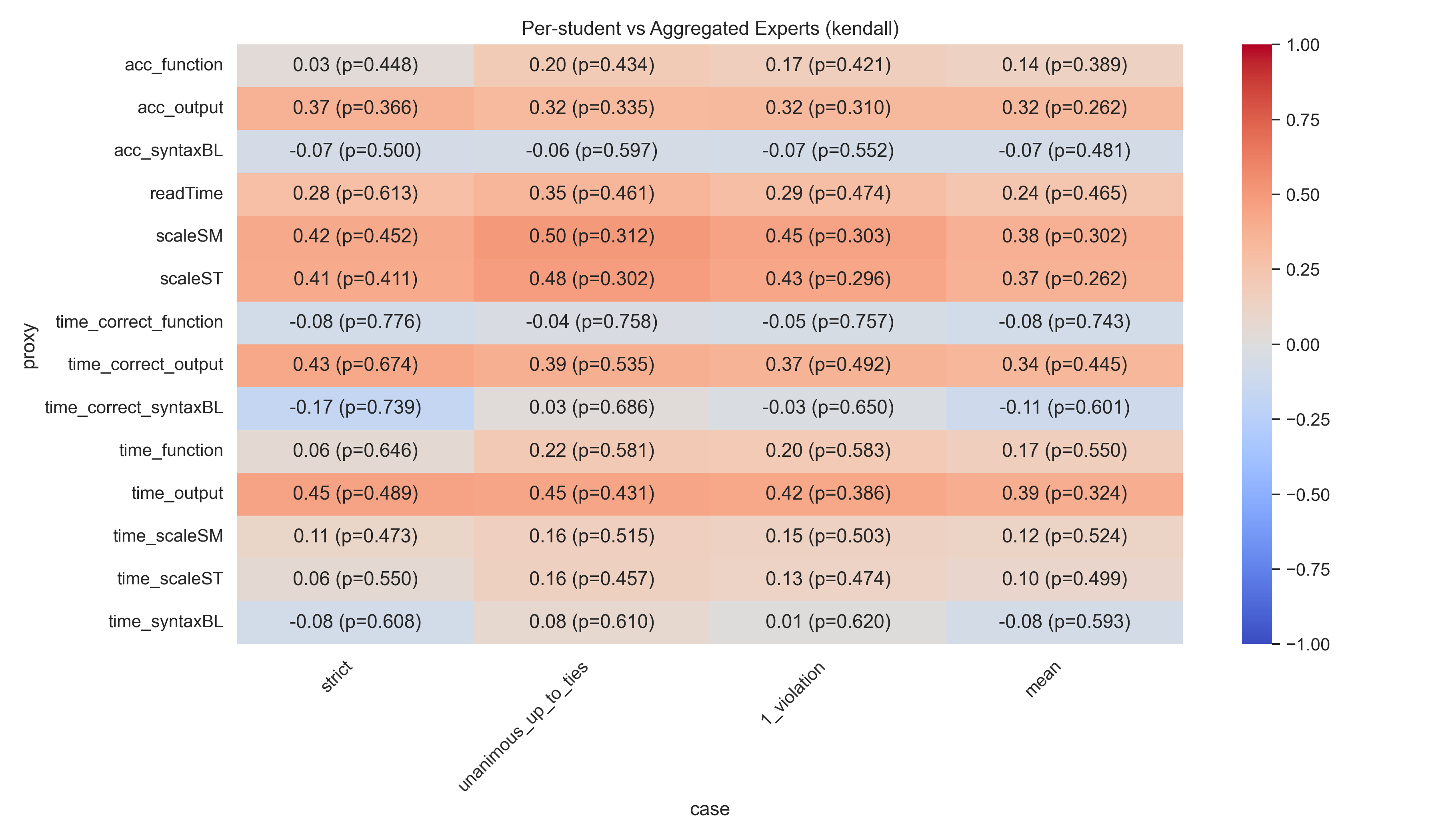}
    \caption{Correlation between per-student proxy rankings and aggregated expert rankings (Kendall).}
    \label{fig:heatmap_perstudent_vs_agg_kendall}
\end{figure}

\begin{figure}[t]
    \centering
    \includegraphics[width=0.9\linewidth]{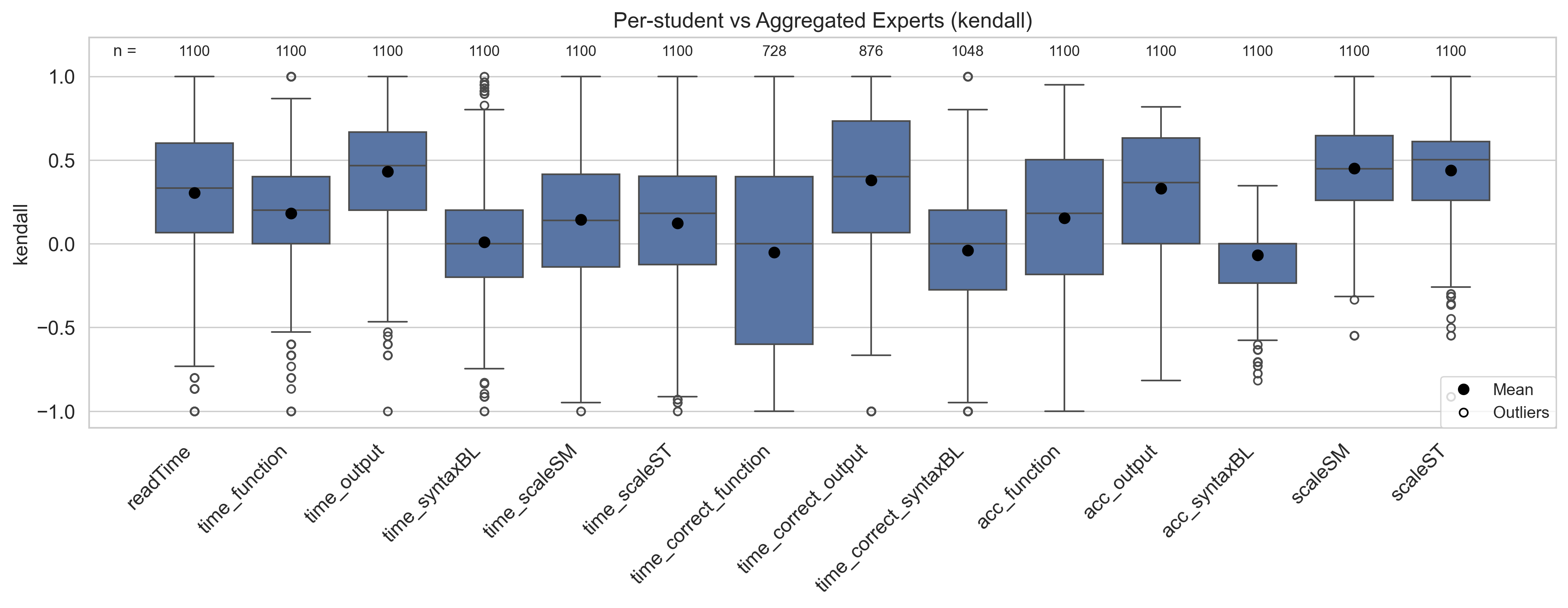}
    \caption{Distribution of correlations between per-student proxy rankings and aggregated expert rankings (Kendall).}
    \label{fig:box_perstudent_vs_agg_kendall}
\end{figure}

Figures~\ref{fig:heatmap_perstudent_vs_agg_kendall} and~\ref{fig:box_perstudent_vs_agg_kendall} show results without aggregating student data. The distributions exhibit higher variability, reflecting noise at the individual level.

\subsection{No Aggregation in Both Sides}
\label{app:no-agg-at-all}

\begin{figure}[t]
    \centering
    \includegraphics[width=0.9\linewidth]{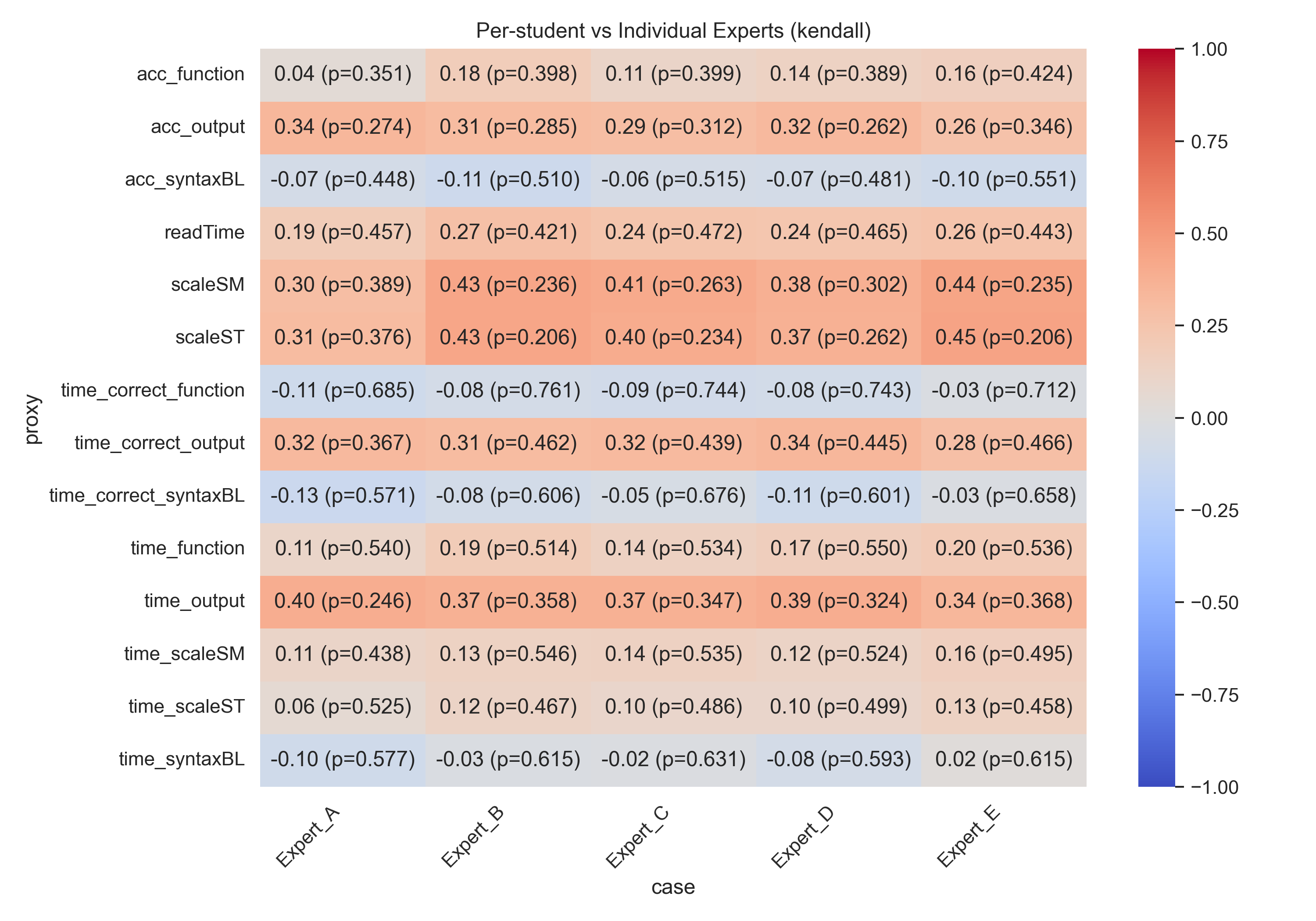}
    \caption{Correlation between per-student proxy rankings and individual expert rankings (Kendall).}
    \label{fig:heatmap_perstudent_vs_expert_kendall}
\end{figure}

\begin{figure}[t]
    \centering
    \includegraphics[width=0.9\linewidth]{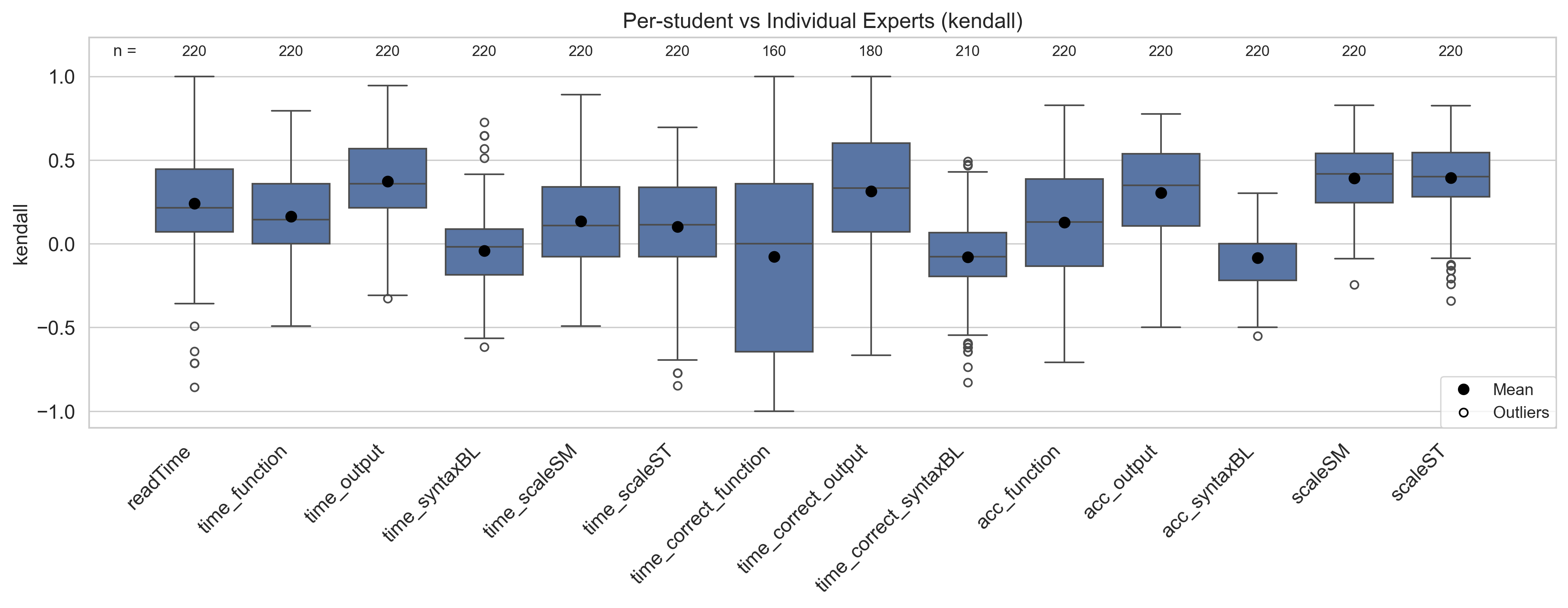}
    \caption{Distribution of correlations between per-student proxy rankings and individual expert rankings (Kendall).}
    \label{fig:box_perstudent_vs_expert_kendall}
\end{figure}

Figures~\ref{fig:heatmap_perstudent_vs_expert_kendall} and~\ref{fig:box_perstudent_vs_expert_kendall} present the fully unaggregated setting. Correlations are more dispersed, but the overall ordering of proxies remains consistent.

\subsection{Impact of Student Characteristics and Fatigue on Code Comprehension Performance}
\label{app:results-background}

\subsubsection{Overview}

We analyze whether student background characteristics and task order influence code comprehension performance in our study. Specifically, we investigate whether self-reported programming experience, object-oriented programming knowledge, background knowledge, and fatigue affect multiple comprehension proxies. Rather than aggregating across question types, we analyze accuracy and time separately for the function, output, and syntax tasks. We also analyze snippet reading time and total time spent on each snippet.

Participants solved eight code comprehension tasks presented in randomized order. For each task, participants first read a code snippet and then answered three comprehension questions: one function question, one output question, and one syntax question. The dataset therefore contains repeated measurements for each participant across multiple snippets and question types.

\subsubsection{Data Structure}

The dataset contains repeated observations for each participant. Each participant solved eight code comprehension tasks, and each task includes three comprehension questions. Therefore, the data can be represented at two levels relevant to the present analysis:

\begin{itemize}

\item \textbf{Question level}: Each observation corresponds to a single comprehension question (function, output, or syntax). This dataset contains up to $44 \times 8 \times 3 = 1056$ observations.

\item \textbf{Snippet level}: Each observation corresponds to a participant--snippet pair. This dataset contains up to $44 \times 8 = 352$ observations.

\end{itemize}

Different statistical models are applied depending on the level at which the outcome variable is defined.

\subsubsection{Measured Outcomes}

We consider the following outcome variables:

\begin{itemize}

\item \textbf{\emph{acc\_function}, \emph{acc\_output}, and \emph{acc\_syntaxBL}}: Binary variables indicating whether the corresponding question was answered correctly. For function questions, correctness is defined using the grading threshold adopted in our study; output and syntax questions use their existing binary correctness labels.

\item \textbf{\emph{time\_function}, \emph{time\_output}, and \emph{time\_syntaxBL}}: Raw time spent answering each question type.

\item \textbf{\emph{time\_correct\_function}, \emph{time\_correct\_output}, and\\
\emph{time\_correct\_syntaxBL}}: Raw time spent answering a question, restricted to cases in which the answer was correct.

\item \textbf{\emph{readTime}}: The time spent reading a snippet before answering questions.

\item \textbf{\emph{snippet\_total\_time}}: The total time spent on a snippet, defined as the sum of the reading time and the times spent answering the three comprehension questions.

\end{itemize}

\subsubsection{Predictor Variables}

We analyze the following participant-level predictors:

\begin{itemize}

\item \textbf{Programming Experience Relative to Peers}: A self-reported measure of programming ability compared to classmates.

\item \textbf{Java Experience (Years)}

\item \textbf{Object-Oriented Experience}: Self-rated experience with object-oriented programming.

\item \textbf{Programming Experience Score}

\item \textbf{Background Knowledge}: A knowledge score derived from responses to eleven multiple-choice familiarity questions about snippets.

\item \textbf{Fatigue}: The position of the snippet in the experiment (1--8), representing task order.

\end{itemize}

All participant-level background predictors were standardized to $z$-scores prior to modeling. Fatigue was operationalized directly as snippet position.

\subsubsection{Statistical Models}

Because participants completed multiple tasks, observations are not independent. We therefore use statistical models that account for repeated measurements.

\paragraph{Accuracy Models}

Accuracy is measured at the question level and is a binary variable indicating whether a comprehension question was answered correctly. We analyze each accuracy proxy separately using logistic regression estimated with clustered standard errors:

\[
\text{accuracy}_{i,j,k} \sim \beta_0 + \beta_1 \text{predictor}_i + \gamma_j
\]

where $i$ indexes participants, $j$ indexes snippets, and $k$ indexes question type. The model includes fixed effects for snippet identity ($\gamma_j$) to control for differences in snippet difficulty, and standard errors are clustered by participant to account for repeated observations.

\paragraph{Time Models}

Question-level time proxies, snippet reading time, and snippet total time are continuous outcomes. We analyze them using linear mixed-effects models:

\[
\text{time}_{i,j,k} \sim \beta_0 + \beta_1 \text{predictor}_i + \gamma_j + u_i
\]

where $u_i$ is a random intercept for participant $i$, capturing individual differences in baseline speed.

\subsubsection{Results and Interpretation}

\Cref{tab:student_factors_final} summarizes the estimated coefficients, standard errors, p-values, and 95\% confidence intervals. Each row in \cref{tab:student_factors_final} corresponds to a separate regression model estimating the effect of a single predictor variable on one comprehension proxy.

For example, the row \emph{Programming Experience Relative to Peers -- acc\_function} estimates the effect of peer-relative programming experience on function-question accuracy, while the row \emph{Fatigue -- time\_output} estimates the effect of snippet order on the time spent answering output questions.

\begin{table}[t]
\scriptsize
\centering
\caption{Impact of participant characteristics and fatigue on comprehension proxies. The table reports results for all estimated models. Coefficients are reported with 95\% confidence intervals. Significant results ($p < 0.05$) are shown in bold.}
\label{tab:student_factors_final}
\begin{tabular}{llllrrr}
\toprule
Analysis & Outcome & Predictor & $\beta$ & SE & $p$ & CI \\
\midrule

\multicolumn{7}{l}{\textbf{PRP}} \\
PRP & acc\_function & PRP & \textbf{0.33} & 0.15 & \textbf{0.024} & [0.04, 0.61] \\
PRP & acc\_output & PRP & \textbf{0.36} & 0.18 & \textbf{0.045} & [0.01, 0.71] \\
PRP & acc\_syntaxBL & PRP & -0.04 & 0.18 & 0.83 & [-0.39, 0.31] \\
PRP & time\_function & PRP & 5.99 & 9.60 & 0.53 & [-12.83, 24.82] \\
PRP & time\_output & PRP & -2.04 & 6.11 & 0.74 & [-14.02, 9.94] \\
PRP & time\_syntaxBL & PRP & -1.05 & 0.76 & 0.16 & [-2.53, 0.43] \\
PRP & time\_correct\_function & PRP & 2.92 & 7.69 & 0.70 & [-12.15, 17.99] \\
PRP & time\_correct\_output & PRP & -5.43 & 6.84 & 0.43 & [-18.84, 7.97] \\
PRP & time\_correct\_syntaxBL & PRP & -1.32 & 0.72 & 0.07 & [-2.73, 0.10] \\
PRP & readTime & PRP & -3.13 & 6.75 & 0.64 & [-16.36, 10.11] \\

\midrule

\multicolumn{7}{l}{\textbf{JavaExp}} \\
JavaExp & acc\_function & JavaExp & -0.12 & 0.18 & 0.51 & [-0.47, 0.24] \\
JavaExp & acc\_output & JavaExp & -0.10 & 0.16 & 0.54 & [-0.42, 0.22] \\
JavaExp & acc\_syntaxBL & JavaExp & -0.43 & 0.25 & 0.09 & [-0.91, 0.06] \\
JavaExp & time\_function & JavaExp & 11.82 & 9.48 & 0.21 & [-6.76, 30.39] \\
JavaExp & time\_output & JavaExp & 7.03 & 6.12 & 0.25 & [-4.97, 19.03] \\
JavaExp & time\_syntaxBL & JavaExp & 0.32 & 0.77 & 0.68 & [-1.20, 1.83] \\
JavaExp & time\_correct\_function & JavaExp & 7.20 & 8.28 & 0.38 & [-9.03, 23.43] \\
JavaExp & time\_correct\_output & JavaExp & 4.58 & 7.07 & 0.52 & [-9.27, 18.43] \\
JavaExp & time\_correct\_syntaxBL & JavaExp & 0.17 & 0.85 & 0.84 & [-1.50, 1.84] \\
JavaExp & readTime & JavaExp & -2.35 & 6.76 & 0.73 & [-15.60, 10.90] \\

\midrule

\multicolumn{7}{l}{\textbf{OOP}} \\
OOP & acc\_function & OOP & \textbf{0.39} & 0.18 & \textbf{0.026} & [0.05, 0.74] \\
OOP & acc\_output & OOP & \textbf{0.45} & 0.20 & \textbf{0.023} & [0.06, 0.84] \\
OOP & acc\_syntaxBL & OOP & 0.03 & 0.20 & 0.89 & [-0.37, 0.42] \\
OOP & time\_function & OOP & 1.50 & 9.65 & 0.88 & [-17.41, 20.40] \\
OOP & time\_output & OOP & -4.06 & 6.11 & 0.51 & [-16.04, 7.92] \\
OOP & time\_syntaxBL & OOP & -1.00 & 0.76 & 0.19 & [-2.49, 0.48] \\
OOP & time\_correct\_function & OOP & -2.81 & 7.83 & 0.72 & [-18.16, 12.55] \\
OOP & time\_correct\_output & OOP & -7.37 & 7.14 & 0.30 & [-21.37, 6.63] \\
OOP & time\_correct\_syntaxBL & OOP & -1.42 & 0.86 & 0.10 & [-3.09, 0.26] \\
OOP & readTime & OOP & -2.11 & 6.76 & 0.76 & [-15.36, 11.15] \\

\midrule

\multicolumn{7}{l}{\textbf{PE}} \\
PE & acc\_function & PE & 0.20 & 0.13 & 0.11 & [-0.05, 0.45] \\
PE & acc\_output & PE & 0.28 & 0.15 & 0.06 & [-0.02, 0.58] \\
PE & acc\_syntaxBL & PE & 0.04 & 0.14 & 0.75 & [-0.23, 0.32] \\
PE & time\_function & PE & -3.21 & 9.64 & 0.74 & [-22.09, 15.68] \\
PE & time\_output & PE & -3.48 & 6.11 & 0.57 & [-15.47, 8.50] \\
PE & time\_syntaxBL & PE & -0.63 & 0.77 & 0.42 & [-2.13, 0.88] \\
PE & time\_correct\_function & PE & -1.85 & 7.54 & 0.81 & [-16.62, 12.92] \\
PE & time\_correct\_output & PE & -0.94 & 6.95 & 0.89 & [-14.56, 12.67] \\
PE & time\_correct\_syntaxBL & PE & -0.50 & 0.85 & 0.56 & [-2.17, 1.17] \\
PE & readTime & PE & -2.10 & 6.76 & 0.76 & [-15.35, 11.15] \\

\midrule

\multicolumn{7}{l}{\textbf{BK}} \\
BK & acc\_function & BK & \textbf{0.35} & 0.14 & \textbf{0.010} & [0.09, 0.62] \\
BK & acc\_output & BK & 0.38 & 0.23 & 0.09 & [-0.06, 0.83] \\
BK & acc\_syntaxBL & BK & 0.03 & 0.14 & 0.86 & [-0.25, 0.30] \\
BK & time\_function & BK & -10.93 & 9.50 & 0.25 & [-29.55, 7.69] \\
BK & time\_output & BK & -11.48 & 6.14 & 0.06 & [-23.51, 0.56] \\
BK & time\_syntaxBL & BK & -1.02 & 0.76 & 0.18 & [-2.50, 0.46] \\
BK & time\_correct\_function & BK & -3.39 & 7.59 & 0.66 & [-18.26, 11.49] \\
BK & time\_correct\_output & BK & -11.30 & 6.87 & 0.10 & [-24.77, 2.17] \\
BK & time\_correct\_syntaxBL & BK & -0.91 & 0.74 & 0.22 & [-2.35, 0.54] \\
BK & readTime & BK & -10.21 & 6.58 & 0.12 & [-23.11, 2.70] \\

\midrule

\multicolumn{7}{l}{\textbf{Fatigue}} \\
Fatigue & acc\_function & Fatigue & 0.00 & 0.05 & 0.99 & [-0.09, 0.09] \\
Fatigue & acc\_output & Fatigue & -0.09 & 0.05 & 0.08 & [-0.20, 0.01] \\
Fatigue & acc\_syntaxBL & Fatigue & 0.10 & 0.09 & 0.28 & [-0.08, 0.27] \\
Fatigue & time\_function & Fatigue & \textbf{-7.95} & 2.04 & \textbf{0.0001} & [-11.95, -3.95] \\
Fatigue & time\_output & Fatigue & \textbf{-7.45} & 1.88 & \textbf{0.0001} & [-11.14, -3.76] \\
Fatigue & time\_syntaxBL & Fatigue & -0.22 & 0.26 & 0.40 & [-0.74, 0.30] \\
Fatigue & time\_correct\_function & Fatigue & -2.90 & 2.25 & 0.20 & [-7.31, 1.51] \\
Fatigue & time\_correct\_output & Fatigue & \textbf{-5.29} & 2.30 & \textbf{0.021} & [-9.79, -0.79] \\
Fatigue & time\_correct\_syntaxBL & Fatigue & \textbf{-0.50} & 0.25 & \textbf{0.043} & [-0.98, -0.02] \\
Fatigue & readTime & Fatigue & \textbf{-5.04} & 1.17 & \textbf{<0.001} & [-7.34, -2.75] \\
Fatigue & snippet\_total\_time & Fatigue & \textbf{-15.62} & 2.67 & \textbf{<0.001} & [-20.86, -10.37] \\

\bottomrule
\end{tabular}
\end{table}

\paragraph{Programming Experience Relative to Peers}

Programming experience relative to peers shows a significant positive association with \emph{acc\_function} and \emph{acc\_output}. Participants who perceive themselves as stronger programmers relative to their peers are more likely to answer function and output questions correctly. In contrast, this predictor does not show a significant relationship with \emph{acc\_syntaxBL} or with any of the time-based proxies.

This pattern suggests that self-perceived programming ability is more closely related to semantic comprehension performance than to syntax-focused comprehension or response speed.

\paragraph{Java Experience (Years)}

Java experience does not show a statistically significant relationship with any of the three accuracy proxies reported in \cref{tab:student_factors_final}. The coefficient for \emph{acc\_syntaxBL} is negative and marginal, but it does not reach conventional significance levels. We therefore do not find evidence that the amount of Java experience, as self-reported by participants, reliably predicts comprehension performance in this study.

\paragraph{Object-Oriented Experience}

Object-oriented programming experience shows a statistically significant positive relationship with both \emph{acc\_function} and \emph{acc\_output}. Participants with stronger self-reported object-oriented experience tend to perform better on these two semantically oriented comprehension tasks. No significant relationship is observed for \emph{acc\_syntaxBL}.

\paragraph{Programming Experience Score}

Programming experience score does not show a statistically significant relationship with any of the reported accuracy proxies. The coefficient for \emph{acc\_output} is positive and close to significance, but the evidence is not strong enough to support a reliable effect.

\paragraph{Background Knowledge}

Background Knowledge shows a statistically significant positive association with \emph{acc\_function}, indicating that participants with stronger background knowledge are more likely to answer function questions correctly. No significant relationship is observed for \emph{acc\_output} or \emph{acc\_syntaxBL}, although the coefficient for \emph{acc\_output} is positive and marginal.

\paragraph{Fatigue Effects}

Fatigue, operationalized as snippet position, does not show statistically significant effects on the accuracy proxies reported in \cref{tab:student_factors_final}. In contrast, fatigue shows strong negative associations with several time-based proxies. Later snippets are associated with lower \emph{time\_function}, lower \emph{time\_output}, lower \emph{time\_correct\_output}, lower \emph{time\_correct\_syntaxBL}, lower \emph{readTime}, and lower \emph{snippet\_total\_time}.

These results indicate that participants become faster as the study progresses, while their accuracy remains relatively stable. This pattern is more consistent with learning, adaptation, or increased familiarity with the study format than with deterioration in comprehension performance over time.

\subsection{Cross-Institutional Consistency Analysis}
\label{app:institutions}

To assess whether our findings generalize across populations, we compare correlation results obtained from students at University~1 ($n = 37$) and University~2 ($n = 7$). Our goal is not to compare individual correlation values in isolation, but to determine whether the \emph{overall pattern of proxy effectiveness} is consistent across institutions.

For each aggregation configuration (i.e., no aggregation/per-student and mean), we construct a vector of correlation values for University~1 and University~2, where each element corresponds to a specific configuration (i.e., a combination of expert aggregation strategies, expert ranking, and proxy). We then measure cross-institutional agreement by computing the Spearman correlation between these vectors. This captures whether proxies that perform well (or poorly) under one population exhibit similar relative performance under the other.

The results show strong agreement across institutions. In particular, the per-student aggregation yields a high correlation ($\rho = 0.78$), indicating that the relative ordering of proxies is largely preserved between University~1 and University~2. Mean aggregation exhibits slightly lower but still substantial agreement ($\rho = 0.68$). These results suggest that the choice of proxy leads to comparable conclusions across both populations.

To quantify the magnitude of differences, we compute the mean absolute difference between corresponding correlation values. The observed differences are moderate, ranging from $\Delta\rho \approx 0.18$ (per-student) to $\Delta\rho \approx 0.25$ (mean), with median differences around $0.17$--$0.20$. This indicates that while individual correlation values may vary due to sampling variability (especially given the smaller University~2 cohort), these variations do not substantially alter the relative behavior of proxies.

Overall, the high vector-level correlations, combined with moderate absolute differences, demonstrate that our findings are robust across institutions. In particular, the per-student aggregation provides the most stable cross-institutional signal, suggesting it is less sensitive to sample size differences and better captures consistent trends in proxy effectiveness.

\begin{table}[t]
\centering
\caption{Cross-institution consistency between University~1 ($n=37$) and University~2 ($n=7$). We report the Spearman correlation between correlation vectors ($\rho$), capturing agreement in proxy rankings, and the mean/median absolute differences ($|\Delta\rho|$), capturing magnitude of variation.}
\label{tab:cross_institution2}
\begin{tabular}{lccc}
\toprule
\textbf{Aggregation} & \textbf{$\rho$ (vector)} & \textbf{Mean $|\Delta\rho|$} & \textbf{Median $|\Delta\rho|$} \\
\midrule
Per-student & 0.78 & 0.18 & 0.17 \\
Mean        & 0.68 & 0.25 & 0.20 \\
\bottomrule
\end{tabular}
\end{table}

\end{document}